\documentclass[11pt,letterpaper]{article}

\usepackage[T1]{fontenc}
\usepackage[a4paper,top=3cm,bottom=2cm,left=3cm,right=3cm,marginparwidth=1.75cm]{geometry}

\usepackage{mathtools}
\usepackage{amsmath,amstext,amsthm,amsfonts,latexsym,amssymb,graphicx}
\usepackage{eurosym}
\usepackage{tabularx}
\usepackage{dsfont}
\usepackage{subcaption} 
\usepackage{graphicx}
\usepackage{algorithm2e,algorithmic}
\usepackage[dvipsnames]{xcolor}
\usepackage{comment,float}

\newtheorem{definition}{Definition}

\newtheorem{proposition}{Proposition}
\newtheorem{theorem}{Theorem}

\newtheorem{assumption}{Assumption}

\newtheorem{fact}{Fact}

\newcommand{\PP}{\mathbb{P}}
\newcommand{\EE}{\mathbb{E}}

\newcommand{\calU}{\mathcal{U}}
\newcommand{\calA}{\mathcal{A}}
\newcommand{\calB}{\mathcal{B}}
\newcommand{\calS}{\mathcal{S}}
\newcommand{\calV}{\mathcal{V}}
\newcommand{\calE}{\mathcal{E}}

\newcommand{\calG}{\mathcal{G}}

\newcommand{\calP}{\mathcal{P}}

\newcommand{\calO}{\mathcal{O}}

\newcommand{\red}{\textcolor{red}}

\DeclarePairedDelimiter\floor{\lfloor}{\rfloor}
\DeclarePairedDelimiter\ceil{\lceil}{\rceil}

\newcolumntype{R}[1]{>{\raggedleft\arraybackslash }b{#1}}
\newcolumntype{L}[1]{>{\raggedright\arraybackslash }b{#1}}
\newcolumntype{C}[1]{>{\centering\arraybackslash }b{#1}}

\newcolumntype{L}{>{\raggedright\arraybackslash}X}

\title{Parity Games}
\author{Richard Combes and Mikael Touati}
\date{}

\begin{document}

	\title{Solving Random Parity Games in Polynomial Time}
	
	

\maketitle

\begin{abstract}
	We consider the problem of solving random parity games. We prove that parity games exibit a phase transition threshold above $d_P$, so that when the degree of the graph that defines the game has a degree $d > d_P$ then there exists a polynomial time algorithm that solves the game with high probability when the number of nodes goes to infinity. We further propose the {\tt SWCP} (Self-Winning Cycles Propagation) algorithm and show that, when the degree is large enough, {\tt SWCP} solves the game. Furthermore, the complexity of {\tt SWCP} is polynomial $O\Big(|\calV|^2 +  |\calV||\calE|\Big)$. The design of {\tt SWCP} is based on the threshold for the appearance of particular types of cycles in the players' respective subgraphs. We further show that non-sparse games can be solved in constant time $O(|\calV|^2)$ with high probability, and emit a conjecture concerning the hardness of the $d=2$ case.
\end{abstract}



\section{Introduction and Contribution}
\subsection{Parity Games}
We consider parity games. Parity games are a family of infinite two player games with complete information whose rules are as follows. The game is played over a directed graph $\calG = (\calV,\calE)$ between two players called ''odd'' (also called player $+1$) and ''even'' (also called player $-1$). Each node $v \in \calV$ has an owner denoted by  $a(v) \in \{-1,+1\}$, and an integer priority $p(v) \in \mathbf{N}$.

The game is played by moving a token on the nodes of the graph sequentially. At time $t=0,1,...$, the token is located at some node of the graph denoted by $v_t \in \calV$, and the player $a(v_t)$ whom owns node $v_t$ selects $v_{t+1}$ amongst the successors of $v_t$, namely $(v_t,v_{t+1}) \in \calE$. The player may choose $v_{t+1}$ based on the whole history $v_0,...,v_{t-1},v_t$. If $v_{t+1}$ is chosen solely as a function of $v_t$ the strategy is called memoryless. In short, all the possible plays $v_0,v_1,...,$ of a given parity game are all the possible directed paths in the directed graph $\calG$.

\begin{figure}[h]
    \centering
    \includegraphics[scale=0.4]{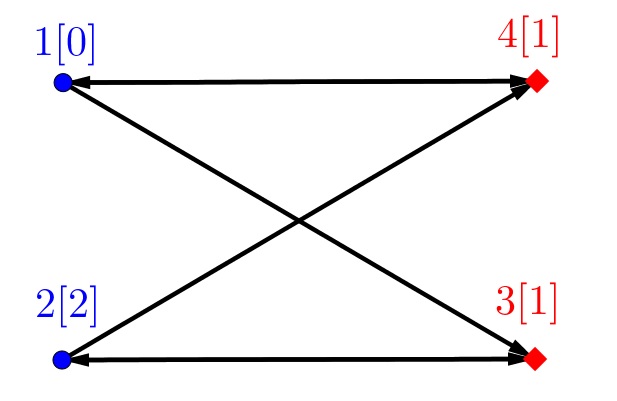}
    \caption{A simple parity game.}
    \label{fig:1}
\end{figure}

Given a play  $v_0,v_1,...$, the player whom wins the game is determined by the following rule: if $\lim\sup_{t \to \infty} p(v_t)$ is odd then player ''odd'' wins, otherwise player ''even'' wins. In short the goal of each player is to make sure that the highest priority that occurs infinitely many times is of her parity. Figure \ref{fig:1} is an example of a simple parity game with four nodes, where the priorities are specified next to the nodes. Nodes owned by player ''even'' and ''odd' are shown in blue and red respectively, and the priority of each node $v$ is indicated as $v [p(v)]$,  Here are two example of plays of this parity game:
\begin{enumerate}
\item Play $1 \to 4 \to 1 \to 4 \to 1 \to 4 \to ...$, so that $\lim\sup_{t \to \infty} p(v_t) = 1$ and player ''odd'' wins. \item Play $1 \to 3 \to 2 \to 3 \to 2 \to 3 \to 2 ...$ so that $\lim\sup_{t \to \infty} p(v_t) = 2$, player ''even'' wins. 
\end{enumerate}
In fact, we can see that, regardless of the starting node $v_0$ player ''even'' can force the token to alternate between nodes $2$ and $3$ after some finite time so that $\lim\sup_{t \to \infty} p(v_t) = 2$, and player ''even'' always wins.


\subsection{Worse Case Complexity of Parity Games}

The main goal is in fact to compute $b(v)$ for all $v \in \calV$, the winner of the game if the game starts at node $v$. We also call $b$ the value of the game. \cite{Mcnaughton1993,Zielonka1998} show that memoryless strategies are optimal, so there is no need to consider non-memoryless strategies to compute its value. Since there are finitely many memoryless strategies, and under any fixed memoryless strategy $v_0,v_1,...$ is periodic after a certain rank, computing $\lim\sup_{t \to \infty} p(v_t)$ for all memoryless strategies yields the value of the game in finite time. 

\begin{proposition}
	Consider a parity game and a starting node $v \in \calV$. Then the game has a winner denoted by $b(v)$, and there exists a winning memoryless strategy. The winner of the game $b(v)$ can be computed in finite time.
\end{proposition}

The natural question is whether or not parity games can be solved in polynomial time. Unfortunately this is an open problem, and so far we only know that this problem is in $QP$, in $NP$ as well as in $coNP$. The literature proposes several quasi-polynomial time algorithms highlighted in the next section, but no polynomial algorithm is known.

\subsection{Worse Case vs Average Complexity}

In this work, we depart from all known work on parity games by studying their average case complexity instead of their worse case complexity. Our motivation is threefold: (i) determining whether or not \emph{typical} instances of parity games may be solved in polynomial time, and what is the influence of the number of nodes and the degree of the game on its difficulty (ii) finding algorithms that potentially perform well on "typical" instances, which are interesting in practice (iii) being able to draw random instances from "easy" or "hard" distributions, which is useful for benchmarking.

Given $|\calV|$ nodes and $|\calE|= d |\calV|$ edges, where $d$ is the degree of a node, a natural distribution over parity games is as follows (see Section~\ref{sec:main_result} for further discussion) :
\begin{itemize}
\item Graph $\calG$ is chosen uniformly from the set of $d$-regular directed graphs with $|\calV|$ nodes i.e. each node has out-going degree $d$.
\item The priority of each node is drawn in an i.i.d. fashion from some distribution over the set of integers 
\item The owner of each node is drawn in an i.i.d. fashion uniformly from $\{-1,+1\}$
\end{itemize}
We will be interested in the large game regime where $|\calV| \to \infty$. We will consider two cases: sparse games where degree $d$ is fixed and $|\calV|$ goes to infinity, and non-sparse games where the degree increases with the number of nodes $d = f(|\calV|)$, where $f$ is some function with $f(|\calV|) \to \infty$ when $|\calV| \to \infty$.

\subsection{Our Contribution}

Our main contribution can be summarized as follows. 

(i) We propose a polynomial-time algorithm called {\tt SWCP} (Self-Winning Cycles Propagation) which runs in $O\Big(|\calV|^2 +  |\calV||\calE|\Big)$ time.

(ii) For sparse games, we prove that there exists a phase transition threshold $d_{P}$ such that, if  $d \ge d_{P}$ then {\tt SWCP} solves the game with arbitrarily high probability as $|\calV|$ grows to infinity. In short, above threshold $d_{P}$, parity games are solvable in polynomial time with high probability.

(iii) As a corollary of our analysis we also prove that, non-sparse parity games can be solved in time $O(|\calV|)$ with high probability using a trivial algorithm, so that sparse parity games are, perhaps surprisingly, the hardest games from the point of view of average case complexity.

(iv) We emit the conjecture that parity games are hard to solve with high probability for $d=2$ and we provide some arguments to justify our intuition. We leave this conjecture as an open problem.


\section{Related Work}

We highlight here some of the relevant related work. We provide more information about related work in appendix, in subsections~\ref{subsec:extended_related_work}, \ref{subsec:bipartite_games}, \ref{subsec:beyond_parity_games}, and  \ref{subsec:mu_calculus}, in particular the links between parity games, other types of games and model checking. We denote by $c = |\{p(v): v \in \calV\} |$ the number of distinct priorities.

\subsection{Worse Case Complexity of Parity Games}

Computing the value of a parity game is known to be in NP $\cap$ Co-NP \cite{Emerson1993,Emerson2001} and even UP $\cap$ co-UP \cite{Jurdzinski1998}. However it is not known to be in $P$, an intriguing status shared by only few problems \cite{Jurdzinski1998}. Recently, \cite{Calude2017} show that the problem is in QP. However whether or not computing the value of a parity game is in $P$ remains an important open problem in computer science.  Computing the value of a parity game also is in PLS, PPAD and CLS \cite{Daskalakis2011}.  Furthremore, if one can compute the value of a parity game in time $O(g(|\calV|,|\calE|,c))$ for some function $g$, then computing memoryless winning strategies for this game can be done in time $O(|\calV| |\calE| g(|\calV|,|\calE|,c))$ by backwards induction.

\subsection{Algorithms for Solving Parity Games}

From an algorithmic perspective, parity games were initially studied as Muller games by \cite{Mcnaughton1993}\cite{Zielonka1998}, showing memoryless determinacy and algorithms to compute the game value. Zielonka's algorithm is exponential \cite{Friedmann2011} and can be implemented in time $O\Big(|\calE|(|\calV|/c)^c\Big)$ \cite{Jurdzinski2000}, performs very well in practice \cite{Friedmann2009}\cite{vanDijk2018} and keeps on inspiring research \cite{Parys2019a}.

These works have been further developed using techniques as progress measures (see \cite{Jurdzinski2017} for an overview of progress measures for parity games) \cite{Jurdzinski2000}, big steps and optimized recursions \cite{Jurdzinski2008}\cite{Jurdzinski2006}\cite{Schewe2007}, strategy improvement \cite{Voge2000}\cite{Schewe2008} and randomization \cite{Petersson2001}\cite{Bjorklund2003}. 

The recent breakthrough~\cite{Calude2017} gives the first known quasi-polynomial algorithm which computes the value of a parity game in time  $O(|\calV|^{\ceil{\ln |\calV|}+6})$  and winning strategies in time $O(|\calV|^{(\ln c)+7}\ln |\calV|)$. 
The algorithm is efficient only when the number of priorities is small, and if  $c<\ln |\calV|$, the time complexity is $O(|\calV|^5)$, subsequently improved to $O(|\calE||\calV|^{2.55})$ in \cite{Gimbert2017}.
The algorithm uses a polylogarithmic-space safety automaton (deciding the winner of a play) and its combination with the original parity game into an easily solvable safety game.

Subsequent contributions \cite{Fearnley2017}\cite{Fearnley2019}\cite{Jurdzinski2017}\cite{Lehtinen2018}\cite{Parys2019b} have refined the analysis and proposed new succinct coding techniques and algorithms.  \cite{Czerwinski2019} shows (also see \cite{Parys2019a,Parys2019b,Jurdzinski2017}) a unified perspective on this line of works within the scope of the separation approach, a technique relying on  separating automata (nondeterministic in \cite{Lehtinen2018}, deterministic for others). 

\cite{Parys2019a} shows a quasi-polynomial quadratic-space modification of Zielonka’s algorithm. \cite{Chatterjee2004,deAlfaro2004} consider stochastic generalizations of parity games and characterize optimal strategies.  From a practical perspective, \cite{vanDijk2018} compares various algorithms (see \cite{Friedmann2009} for more tests), showing that Zielonka's algorithm \cite{Zielonka1998} and priority promotion \cite{Benerecetti2018} perform efficiently, leaving room for new practically efficient quasi-polynomial algorithms.

\subsection{Average Case Complexity and Phase Transition Phenomena}

There has been a large body of work linking average case complexity and phase transition  phenomena~\cite{Mezard}. Typically one considers an optimization problem over a large graph with given degree, and based on the degree $d$, the optimization problem features phase transitions from solvable in polynomial time  to difficult to solve, or even having no feasible solutions. Problems that feature phase transitions include K-SAT and 3-SAT~\cite{Mezard}, community detection in the stochastic block model~\cite{Bordenave}, planted clique~\cite{Gamarnik} and several others.


\section{Main Result}\label{sec:main_result}

\subsection{Assumptions and Notation}

	Before stating the main result, we recall some notation and state the assumptions made on the distribution over the set of parity games considered. We recall that a parity game is a tuple $(\calG,a,p)$, where $\calG = (\calV,\calE)$ is a directed graph, $a(v) \in \{-1,+1\}$ for $v \in \calV$ indicates the player whom owns node $v$, and $p(v) \in \mathbf{N}$ for $v \in \calV$ is a positive integer indicating the priority of node $v$. The degree  of the game is denoted by $d \ge 1$. We denote by $b(v) \in \{-1,+1\}$ the player whom wins the game starting at node $v$ and the goal is to compute $b$.  The statistical assumptions are described below. In particular, we assume that the game is not biaised towards any of the players, in the sense that, on average, each player owns half of the nodes, and the number of priorities that are odd and even are the same. This is to avoid biases in the sense that, if player ''odd'' (respectively ''even'') owns most of the nodes, or if most of the parities are odd (respectively ''even''), then one player has an overwhelming chance of winning the game. We leave the study of the biased case for further work.

\begin{assumption}\label{ass:random_graph}
	Graph $\calG$ is drawn uniformly at random from the set of directed graphs where each node has out-going degree $d$. The ownership of each node $a(v)$ is drawn in an i.i.d. fashion from the uniform distribution over $\{-1,1\}$. The priority of each node $p(v)$ is drawn in an i.i.d. fashion from a distribution $\calP$ over $\mathbf{N}$. Distribution ${\cal D}$ is balanced in the sense that:
	$$
		\PP(p(v) \text{ is even} ) = \PP(p(v) \text{ is odd} ) = {1 \over 2}.
	$$
\end{assumption}

We will consider the large graph regime where $\calV\to \infty$. Given some event $E$, we say that event $E$ occurs with high probability (w.h.p.) if and only if $\PP(E) \to 1$ when $|\calV| \to \infty$.

\subsection{Main Theorem}
	
	Our main result is that there exists a threshold $d_P \ge 1$ such that, in the sparse graph regime, there exists a polynomial time algorithm with complexity $O\Big(|\calV|^2 +  |\calV||\calE|\Big)$ which outputs the winner of the game $b(v)$ w.h.p. In short, $d_P$ is a phase transition above which parity games can be solved in polynomial time w.h.p. We call this algorithm {\tt SWCP} (Self-Winning Cycle Propagation) and we describe it in full details below. 
	
	This result is somewhat surprising in the sense that, for any fixed degree $d > 1$, no known polynomial algorithm exists, and suggests that, while there may exist instances which are difficult to solve, those instances are not "typical". Since "typical" instances are solvable in polynomial time as long as the degree $d$ is not too small, {\tt SWCP} seems like an interesting solution for solving parity games in practice. This also implies that, in order to obtain hard instances of parity games to benchmark algorithms, sampling sparse parity games with degree below the threshold $d < d_P$. 

We also show that non-sparse parity games are even easier to solve, since they can be solved in time $O(|\calV|)$ by a trivial algorithm. This is also a bit surprising since the total number of strategies is $d^{|\cal V|}$, so that intuitively one would think that solving parity games gets harder as $d$ increases due to having a larger number of solutions to consider.

\begin{assumption}\label{ass:threshold}
The graph degree satisfies:
	$$
		d \eta\Big(d-1,{1 \over 4}\Big) < 1
	$$
	where $\eta(d,q)$ is the extinction probability of a branching process with offpsring distribution Binomial$(d,q)$, i.e. the smallest solution to equation:
	$$
		(1-q + q \eta)^{d} = \eta 
	$$
	Equivalently we have:
	$$
	    d({3 \over 4} + {1 \over 4 d})^{d-1} \le 1
	$$
\end{assumption}

\begin{theorem}\label{th:main_result}
	Consider assumption~\ref{ass:random_graph}. Then there exists a threshold $d_P \ge 1$ such that there exists a polynomial time algorithm which solves parity games w.h.p. if $d>d_P$. 

	 Algorithm {\tt SWCP} can be implemented in time $O\Big(|\calV|^2 +  |\calV||\calE|\Big)$, and if assumption~\ref{ass:threshold} holds as well, it solves parity games w.h.p.
\end{theorem}

Both the rationale for the design of the {\tt SWCP} algorithm as well as the proof of Theorem~\ref{th:main_result} are deeply rooted in the link between branching processes and the existence of particular types of cycles in the subgraph of nodes owned by each player. The proof of Theorem~\ref{th:main_result} is fully presented in section~\ref{sec:analysis}. A reminder of branching processes is presented in appendix, section \ref{subsec:branching_reminder}.

\subsection{Self-Winning Cycles and Nodes}

The {\tt SWCP} algorithm relies on the concept of self-winning cycles and self-winning nodes, which forms the cornerstone of its design. For $i={-1,+1}$ we denote by ${\calG}_{i}$ the subgraph formed by the nodes owned by player $i$. Furthermore, we call~\emph{self-winning} a cycle in subgraph ${\calG}_{i}$ whose maximal parity is even (if $i = -1$) or odd (if $i = +1$). We further say that that node $v \in \calV$ is a self-winning node if it is included in a self-winning cycle.

\begin{fact}
	If node $v \in \calV$ is a self-winning node then the game is won by the player whom owns node $v$ so that $b(v) = a(i)$.
\end{fact}

The reason for this fact is simple: if the game starts in node $v$ owned by player $a(v) = i$, then player $i$ can force the token to cycle infinitely, in a cycle whose highest priority has the parity that guarantees a win for player $i$. We speak of "self-winning" nodes and cycles, because they allow player $i$ to win by forcing the token to stay trapped on the set of nodes that belongs to her, without letting the opponent take any decision.

\begin{figure}
    \centering
    \includegraphics[scale=0.4]{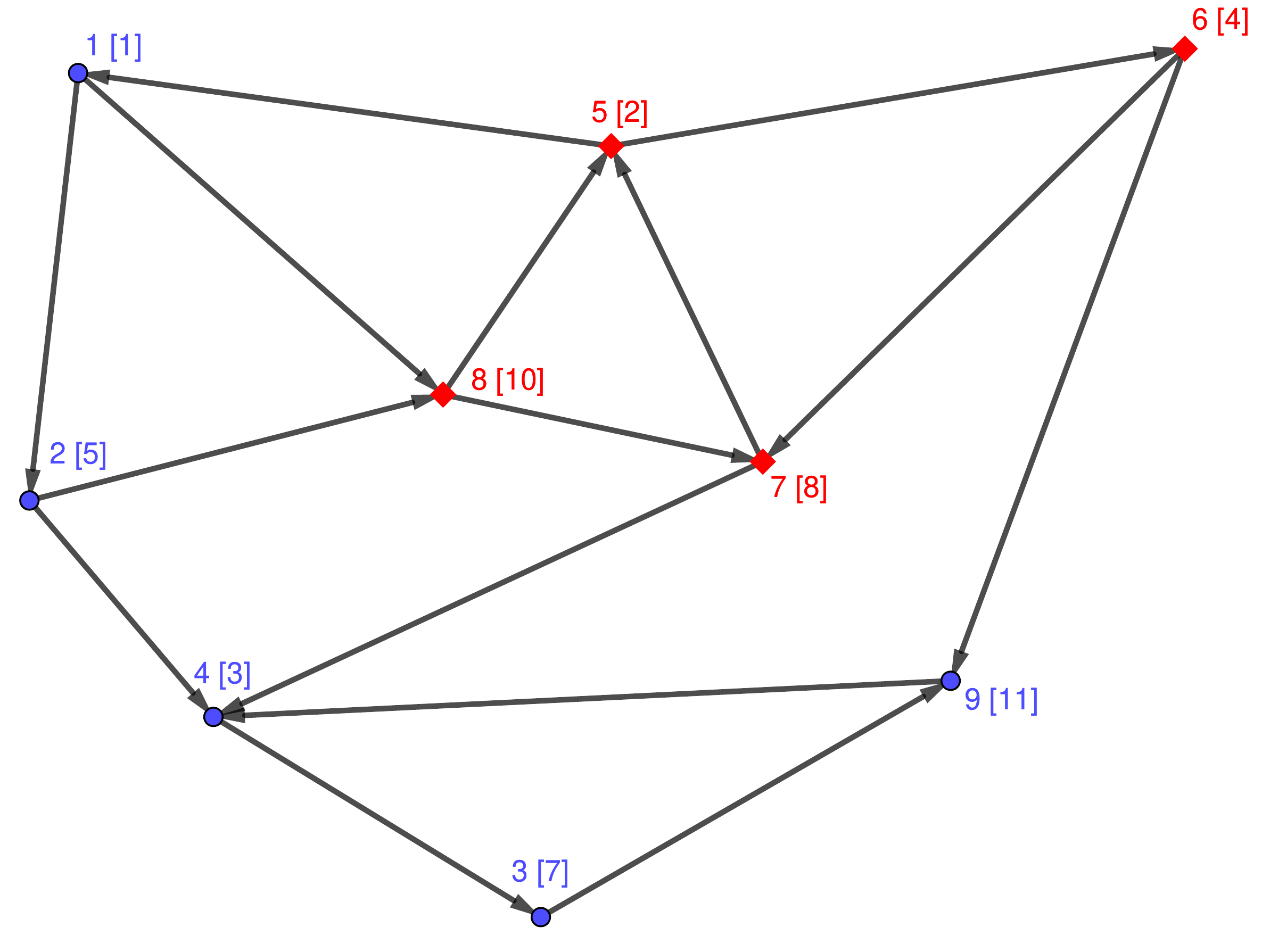}
    \caption{A parity game with several easy cycles.}
    \label{fig:2}
\end{figure}

On figure~\ref{fig:2} we display an example of a parity game. There is one easy cycle for player $1$ which is $4 \to 3 \to 9 \to 4$, and nodes $1$ and $2$ lead to this easy cycle. Therefore, if the game starts at nodes $1,2,3,4,9$ then player $1$ wins. There is one easy cycle for player $2$ which is $5 \to 6 \to 7 \to 5$ and node $8$ leads to this cycle. Therefore, if the game starts at nodes $5,6,7,8$ then player 2 wins. This is a case in which finding all easy cycles of the game enables us to completely solve it.

\subsection{Dynamic Programming}

The {\tt SWCP} algorithm further uses dynamic programming as a building block. As in all optimal control problems, in order to find the optimal strategy in the current state (i.e. the current location of the token), it is sufficient to know the optimal strategy for all successors. 

\begin{fact}
	For any node $v \in \calV$ we have $a(v) b(v) =  \max_{v': (v,v') \in \calE} \{ a(v) b(v') \}$.
\end{fact}

Indeed, consider node $v \in \calV$ owned by player $a(v) = i$. Then there exists a winning strategy for player $i$ starting at $v$ if and only if there exists $v'\in \calE$ a successor of $v$ (i.e. $(v,v') \in E$) and there exists a winning strategy for player $i$ starting at $v'$. Another way of saying this is that player $i$ has a winning strategy from $v$ unless its opponent has a winning strategy starting from all successors of $v$. This is interesting because it implies that if we have already determined the value of a subset of nodes, we can infer the value of their predecessors by using backwards induction.

\subsection{The {\tt SWCP} Algorithm and Complexity}

\begin{definition} The {\tt SWCP} algorithm first finds all self-winning nodes. Once this is done it attempts to compute the value of each node by dynamic programming. 
\end{definition}

\begin{proposition} 
    The {\tt SWCP} algorithm can be implemented in time $O\Big(|\calV|^2 +  |\calV||\calE|\Big)$.
\end{proposition} 

The {\tt SWCP} can be implemented in time  $O\Big(|\calV|^2 +  |\calV||\calE|\Big)$ as follows. The pseudo code of {\tt SWCP} is presented as algorithm~\ref{algo:swcp} for completeness. The subroutine {\tt DFS} is depth first search, so that {\tt DFS}$(v,\calV,\calE)$ outputs the set of  all nodes that can be reached from $v$ in graph $\calG = (\calV,\calE)$, and runs in time $O\Big(|\calV| + |\calE|\Big)$.

{\bf Finding Self-Winning Nodes} Denote by ${\bf par}$ the function that outputs the winning player for a given priority, so that ${\bf par}(p) = +1$ if $p$ is odd and ${\bf par}(p) = -1$ if $p$ is even. Assume that node $v \in \calV$ is included in a self-winning cycle. Denote by $\bar{v}$ the node of this cycle with maximal parity. We must have ${\bf par}(p(\bar{v})) = a(\bar{v})$. and this cycle is a cycle in the subgraph $\calG' =(\calV',\calE)$ of nodes which all belong to player $a(\bar{v})$ and whose parity is lesser or equal to that of $\bar{v}$, i.e.
$$
    \calV' = \{v' \in \calV: a(v') = a(\bar{v}), p(v') \le p(\bar{v}) \}.
$$
Given $\bar{v}$ and $\calG'$, finding the set of nodes $v$ that are included in a cycle in $\calG'$ that contains  $\bar{v}$ can be done by computing the set of all nodes that can be reached from  $\bar{v}$ in $\calG'$ (using depth-first search), computing the set of all the nodes from which we can reach $\bar{v}$ in $\calG'$ (using depth-first search once again), and taking the intersection of those two sets. Repeating this procedure for all $\bar{v} \in \calV$ such that ${\bf par}(p(\bar{v})) = a(\bar{v})$ yields all the self winning nodes.

Since depth-first search can be done in time $O(|\calV| + |\calE|)$, and that we must perform it twice for each value of $\bar{v} \in \calV$, the complexity of finding all self-winning nodes with this procedure is $O\Big(|\calV|^2 +  |\calV||\calE|\Big)$.

{\bf Dynamic programming} Once the self-winning nodes have been found, we can apply dynamic programming. For all nodes in $v \in \calV$ set $e(v) \gets a(v)$ if $v$ is self-winning and $e(v) \gets 0$ otherwise. Then loop $|\calV|$ times over the nodes $v \in \calV$ in a round robin fashion, and each time one finds a node $v$ such that $e(v) = 0$ and either (i) for all $(v,v') \in E$, $e(v) \in \{-1,1\}$, or (ii) there exists $v'$ such that $(v,v') \in E$ and $a(v) e(v') = 1$ then set:
$$
e(v) \gets a(v) \max_{v': (v,v') \in \calE} \{ a(v) e(v') \}
$$
It is noted that when either (i) or (ii) is true, one can always compute the value of $\max_{v': (v,v') \in \calE} \{ a(v) e(v') \}$ even if $e(v') = 0$ for some $v'$ with $(v,v') \in E$, so that we have enough information to determine the value of node $v$. The dynamic programming step hence requires time $O(|\calV||\calE|)$, and the ${\tt SWCP}$ algorithm indeed runs in time $O\Big(|\calV|^2 +  |\calV||\calE|\Big)$.

\vspace{1cm}

\begin{algorithm}[H]
 \KwData{A parity game $(\calG,p,a)$}
 \KwResult{Game value $e$}
 \tcc{Initialization}
  \For{$v \in \calV$}{$e(v) \gets 0$ \;}
  $\calE^\top \gets \{(v,v') \in \calV \times \calV: (v',v) \in \calE \}$\tcc*[r]{Edges for the inverted graph}
  
  \tcc{Computation of self-winning nodes}
 \For{$\bar{v} \in \calV$}{
    \If{$a(\bar{v}) = {\bf par}(p(\bar{v}))$}{
    $\calV' \gets \{v' \in \calV: a(v') = a(\bar{v}), p(v') \le p(\bar{v}) \}$  \;
    $\calA \gets {\tt DFS}(\bar{v},\calV',\calE )$\tcc*[r]{Nodes reachable from $\bar{v}$}
    $\calB \gets {\tt DFS}(\bar{v},\calV',\calE^\top )$\tcc*[r]{Nodes from which we can reach $\bar{v}$}
    \For{$v \in \calA  \cap \calB$}{
        $e(v) \gets a(v)$\tcc*[r]{Self-winning nodes}
    }
    }
 }
  \tcc{Dynamic programming using backwards induction}
 \For{$t=1,...,|\calV|$}
 {
    \For{$v \in \calV$}{
        \uIf{$\exists v': (v,v') \in \calE, a(v)e(v') = 1$}{$e(v) \gets a(v)$\tcc*[r]{Node wins for $a(v)$}}\uElseIf{$\forall v': (v,v') \in \calE, a(v)e(v') = 1$}{$e(v) \gets -a(v)$\tcc*[r]{Node loses for $a(v)$}}\uElse{$e(v) \gets 0$\tcc*[r]{Value of node $v$ cannot be decided for now}}
    }
}
 \caption{The Self-Winning Cycles Algorithm}
 \label{algo:swcp}
\end{algorithm}

\subsection{Additional Results}

We have previously considered the case of sparse games with degree $d \ge d_P$. We now consider the case of sparse games with degree $d=1$ and the case of non sparse games, where $d = f(|\calV|) \to \infty$ when $|\calV| \to \infty$. In both cases the value of the game can be computed in polynomial time using a trivial algorithm. The proof of theorem~\ref{th:non_sparse} is presented in appendix.

\begin{proposition}
    Consider any parity game with degree $d=1$. Then, for all $v$, the value of the game can be computed in time $O(|\calV|^2)$.
\end{proposition}

If the degree of the game is $d=1$, then there exists only one possible play of the game starting at $v \in \calV$. Let $v = v_0 \to v_1 \to ...$ denote this play. This play is the union of a path and a cycle, and computing the largest parity on this cycle yields the winner of the game. For a given $v$ this can be done in time $O(|\calV|)$ simply by enumerating the successors of $v$ and stopping whenever any node has been seen twice. Doing this for all $v$ takes time $O(|\calV|^2)$ to compute the value of the game. 

\begin{theorem}\label{th:non_sparse}
	Consider assumption~\ref{ass:random_graph}. Consider a parity game with degree $d = f(|\calV|) \to \infty$
	when $|\calV| \to \infty$. Then for all $v \in \calV$ we have:
	$$
	  \PP( b(v)= a(v) )  \to 1 \text{ when } |\calV| \to \infty
	$$
	So the player whom owns the starting node always wins the game, and the value of the game can by computed with high probability in time $O(|\calV|)$ by outputting $a$. 
\end{theorem}

\subsection{The $d=2$ regime and a conjecture}

From theorem~\ref{th:main_result} we know that sparse games with large enough degree $d \ge d_P$ can be solved in polynomial time. We also know that games with degree $d=1$ can be solved in polynomial time. Therefore a natural question would be whether or not one can solve the $d=2$ case in a polynomial time, and we emit the conjecture that this is not the case. This is justified by the fact that, when $d=2$ the probability that a node $v$ is self-winning vanishes. Therefore one cannot hope to obtain the value of a fraction of the nodes by considering self-play. The proof of proposition \ref{prop:d2} is given in appendix.

\begin{proposition}\label{prop:d2}
	Consider assumption~\ref{ass:random_graph}. Consider a parity game with degree $d = 2$. Then for all $v \in \calV$ we have:
	$$
	  \PP( v \text{ is self-winning} )  \to 0 \text{ when } |\calV| \to \infty
	$$
\end{proposition}


\section{Analysis}\label{sec:analysis}

In this section we provide the proof of our main result which is Theorem~\ref{th:main_result}. We first state some preliminary technical results in subsection~\ref{subsec:preliminary}. In subsection~\ref{subsec:dynamic_programming} we provide a necessary condition for the value of a node to be computable by backwards induction. We upper bound the probability that a particular node $v$ does not lead to a self-winning cycle by using a branching process argument in subsection~\ref{subsec:branching}. Subsection~\ref{subsec:puttingit} completes the proof of Theorem~\ref{th:main_result}.

\subsection{Preliminary Results}\label{subsec:preliminary}

{\bf Exploration Process of a Graph} Given a directed graph $\calG = (\calV,\calE)$, and a subset of nodes $\calA \subset \calV$ the exploration process of $\calG$ with initial set $\calA$ is the following iterative process with, at time $t$, $\calA_t$ set of active nodes, $\calS_t$ the set of explored nodes and $\calU_t =  \calV \setminus (\calA_t \cup \calS_t)$ set of unseen nodes. Initially $\calA_0 = \calA$, $\calS_0 = \emptyset$ and $\calU_0 = \calV \setminus \calA_0$. At time $t$ (i) an active node $v \in \calA_t$ is selected and is considered explored $\calS_{t+1} = \calS_{t} \cup \{v\}$ and (ii) its unseen successors become active $\calA_{t+1} = \calA_{t} \cup \{v' \in \calU_t: (v,v') \in \calE\} \setminus \{v\}$.
\\

\noindent 
{\bf Sampling With and Without Replacement} We recall a basic fact about sampling with and without replacement and stochastic ordering.
\begin{definition}
We say that $X \le Y$ in the strong stochastic order if and only if $\PP(X \le z) \le \PP(Y \le z)$ for all $z$.
\end{definition}
\begin{fact}\label{fact:hypergeo}
Consider $Y \sim $ Hypergeometric$(N,K,d)$, $X \sim $ Binomial$(d,{K \over N})$ and \newline $Z \sim $ Binomial$(d,{K - d \over N})$. Then $Z \le Y \le X$ in the strong stochastic order.
\end{fact}
Indeed, consider an urn with $K$ black balls and $N-K$ red balls, $Y$ is the number of black balls obtained by drawing $d$ times without replacement while $X$ is the number of black balls obtained by drawing $d$ times with replacement, and $Z$ is the number of black balls obtained by first replacing $d$ black balls by red balls and subsequently drawing $d$ times from the urn without replacement. Therefore $Z \le Y \le X$.
\\

\noindent {\bf Locally Tree Likeness} Consider $v \in \calV$ and $h \ge 0$, and define ${\cal T}_h(v)$ the set of descendants of $v$ within distance $h$ or less in $\calG$. Namely, $v' \in {\cal T}_h(v)$ if and only if there exists a path in $\calG$ of length less than $h$ from $v$ to $v'$. We have that ${\cal T}_h(v)$ is a tree with high probability in proposition~\ref{prop:locally}. The proof is omitted and follows from showing that the expected number of cycles in subgraph ${\cal T}_{h_{|\calV|}}(v)$ vanishes (see for instance \cite{Hofstad2016}).

\begin{proposition}\label{prop:locally}
	Consider $h_{|\calV|} = o(\ln |\calV|)$, then
	$$
		\PP( {\cal T}_{h_{|\calV|} }(v) \text{ is a tree} ) \to 1 \text{ when } |\calV| \to \infty
	$$
\end{proposition}

\subsection{Dynamic programming}\label{subsec:dynamic_programming}

For a node $v \in \calV$, we write that $\ell(v)=a(v)$ if there exists a path in $\calG_{a(v)}$ from $v$ to a self-winning node and $\ell(v) = 0$ otherwise. It is noted that $\ell(v) = a(v)$ implies $b(v) = a(v)$, since starting from node $v$, player $a(v)$ can force the game to reach a self-winning node, and wins the game.

Consider ${\cal T}_h(v)$ the set of descendants of $v$ within distance $h$ and consider applying dynamic programming to ${\cal T}_h(v)$ (ignoring all other nodes) in order to retrieve $b(v)$ the value of $v$. If this is possible we say that $v$ is decidable at height $h$. It is noticed 
\begin{proposition}
	Assume that ${\cal T}_h(v)$ is a tree, and that any path $v \to v_1 \to ... \to v_h$ in ${\cal T}_h(v)$ contains a node $v_j$, $j \in \{1,...,h\}$ with $\ell(v_j) \ne 0$ Then $v$ is decidable at height $h$.
\end{proposition}
Indeed, consider ${\cal T}'_h(v)$ the tree obtained by removing all descendants of nodes with $\ell(v') \ne 0$. Then we have a tree ${\cal T}'_h$ where the value of all leaves are known. Applying dynamic programming ${\cal T}'_h$ immediately yields $b(v)$.

\begin{figure}[H]
    \centering
    \includegraphics[scale=0.5]{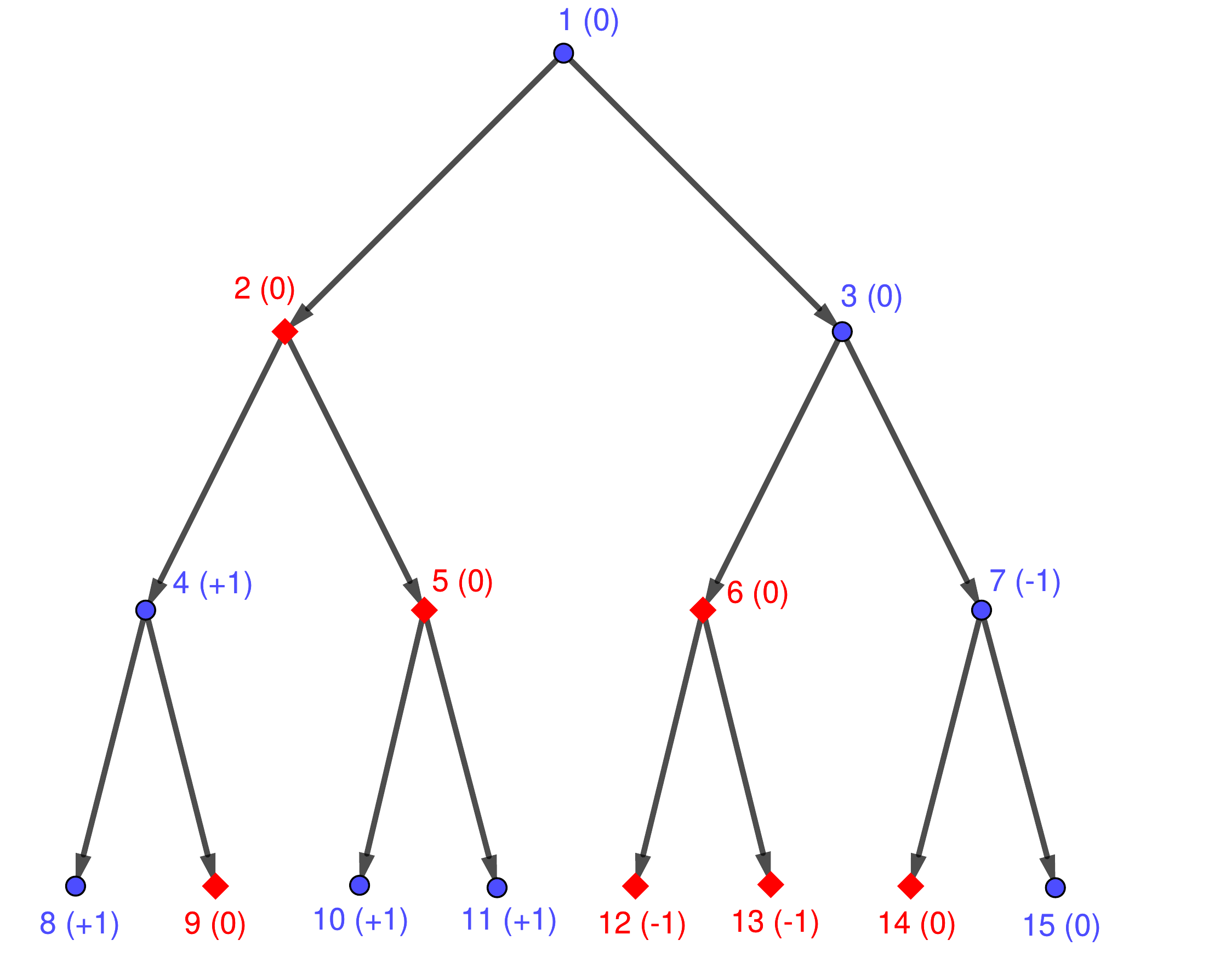}
    \caption{Successors of a given node and markings.}
    \label{fig:3}
\end{figure}

\begin{figure}[H]
    \centering
    \includegraphics[scale=0.5]{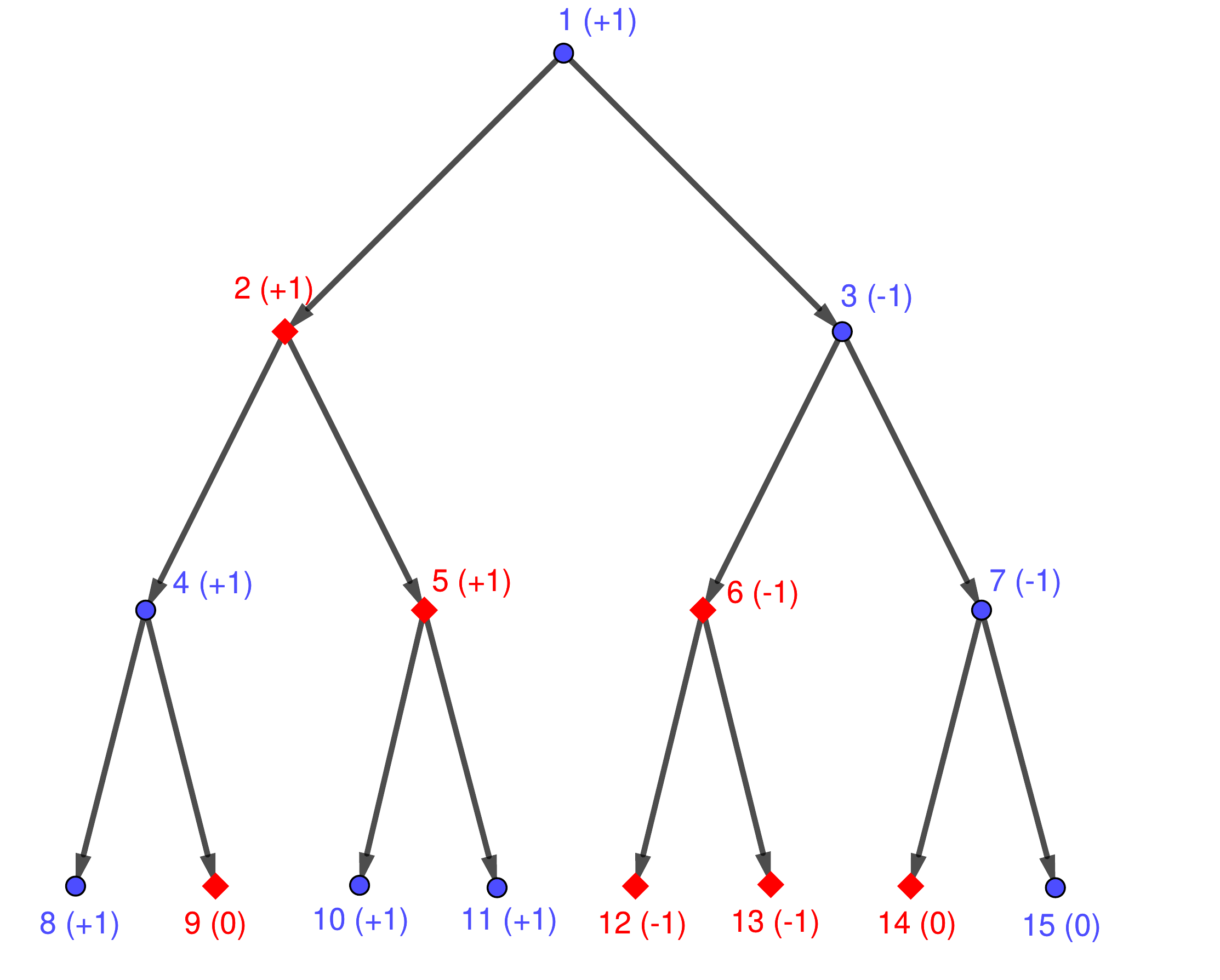}
    \caption{Deciding the value of the game using dynamic programming.}
    \label{fig:4}
\end{figure}

On figure~\ref{fig:3} we display the successors of a given node $v$ at height at most $4$. The  marking of node $v$ is displayed next to it in the format $v (b(v))$. We notice that the initial node is decidable at height $4$, in the sense that all of its successors possess at least one ancestor with $\ell \ne 0$. This means that we can decide the value of node $v$, by using dynamic programming, and the result of dynamic programming is displayed in figure~\ref{fig:4}.


\subsection{Branching, Exploration and apparition of Self-Winning Cycles }\label{subsec:branching}
\noindent  {\bf Branching Processes} Consider the following process defined by $S_0 = k$ and:
\begin{align*}
	S_{t+1} = \begin{cases} S_t - 1 + Z_t \;,\; S_t \ge 1 \\ 0 \text{ otherwise} \end{cases}.
\end{align*}
assume that, for all $1 \le t \le T$, knowing $S_t$, the distribution of $Z_t$ is lower bounded (in the sense of strong stochastic ordering) by the Binomial$(d,q)$ distribution.

Consider an alternate process defined by $S'_0 = k$ and:
\begin{align*}
	S'_{t+1} = \begin{cases} S'_t - 1 + Z'_t , S'_t \ge 1 \\ 0 \text{ otherwise} \end{cases}.
\end{align*}
where $(Z_t')_t$ is i.i.d. with distribution $Z_t \sim$ Binomial$(d,q)$.

Then by a straightforward comparison argument $S_t \ge S'_t$ in the strong stochastic order, so that:
$$
	\PP(S_T = 0) \le \PP(S'_T = 0) \le \lim_{T \to \infty} \PP( S'_T = 0). 
$$
Furthermore, the right hand side is exactly the extinction probability of a branching process with starting population $k$ with progeny distribution Binomial$(d,q)$ so that
$$
	\lim_{n \to \infty} \PP( S'_n = 0) = \eta(d,q)^{k}
$$
where $\eta(d,q)$ is the unique solution to 
$$
	\eta  = \EE(\eta^{Z_t}) = (1 - q + q \eta)^d.
$$

\noindent {\bf Exploration Processes} Consider the exploration process of a directed graph with $|\calV|$ nodes where each node has a number of successors distributed as Binomial$(d,q)$. The process is defined by $S_0 = k$ and:
\begin{align*}
	S_{t+1} = \begin{cases} S_t - 1 + Z_t , S_t \ge 1 \\ 0 \text{ otherwise} \end{cases}.
\end{align*}
Now it is noted that $Z_t$ is the number of successors of the currently explored node that have not been seen yet. Denote by $K_t$ the number of neighbours. $Z_t$ can written as $Z_t \sim $ Hypergeometric$(|\calV|,U_t,K_t)$ with $U_t$ the number of unseen nodes. 

Consider $\varepsilon > 0$, $1 \le t \le T-1$ with $T = {\varepsilon \over d} |\calV|$. Since at most $d$ nodes are seen at each step of the exploration $U_t - d \ge |\calV| - d (t+1) \ge |\calV| - d T = (1-\varepsilon)|\calV|$ and:
$$
	\text{Hypergeometric}(|\calV|,U_t,K_t) \ge \text{Hypergeometric}(|\calV|,|\calV|(1-\varepsilon)+d,K_t) \ge \text{Binomial}(K_t,1 - \varepsilon)
$$
Furthermore, $K_t \sim$Binomial$(d,q)$ so that $Z_t \ge $ Binomial$(d, q(1 - \varepsilon))$.

By the previous comparison argument, we immediately get:
$$
	\PP(S_{T-1} = 0) \le \eta(d,q(1-\varepsilon))^{k} 
$$

\noindent {\bf Apparition of Cycles } Consider the previous exploration process and denote by $N_t$ the number of cycles in the subgraph formed by considering only the nodes that are not inactive. Assume that $S_t > 0$ and denote by $v$ the node explored at time $t$. If one of the successors of $v$ is one of its ancestors (there are at least ${\ln (t/k) \over \ln d}$ of them),  then a cycle appears, so that:
$$
	\PP(N_{t+1} = 0, S_{t+1} > 0) \le  \left(1 - {\ln (t/k) \over |\calV| \ln d}\right)^{d} \PP(N_{t} = 0, S_{t} > 0) 
$$
By recursion:
$$
	\PP(N_{t} = 0, S_{t} > 0) \le \PP(N_{0} = 0, S_{0} > 0) \prod_{s=1}^t \left(1 - {\ln (s/k) \over |\calV| \ln d}\right)^{d} = \prod_{s=1}^t \left(1 - {\ln (s/k) \over |\calV| \ln d}\right)^{d} 
$$
Furthermore:
\begin{align*}
	\prod_{s=1}^t \left(1 - {\ln (s/k) \over |\calV| \ln d}\right)^{d} &\le \prod_{s=t/2}^t \left(1 - {\ln (s/k) \over |\calV| \ln d}\right)^{d}
	&\le \left(1 - {\ln (t/(2 k)) \over |\calV| \ln d}\right)^{d t / 2}
	&\le \exp\left(- {d t \ln (t/(2 k)) \over 2 |\calV| \ln d}\right).
\end{align*}
In particular, for $T = \varepsilon |\calV|/d$:
\begin{align*}
	\PP(N_{T-1} = 0) &= \PP(N_{T-1} = 0, S_{T-1} = 0) + \PP(N_{T-1} = 0, S_{T-1} > 0) \\
	 &\le \PP(S_{T-1} = 0) + \exp\left(- {d (T-1) \ln ((T-1)/(2 k)) \over 2 |\calV| \ln d}\right) \\
	 &\le \eta(d,q(1-\varepsilon))^{k}  + \exp\left(- {d (\varepsilon |\calV|/d-1) \ln ((\varepsilon |\calV|/d-1)/(2 k)) \over 2 |\calV| \ln d}\right)
\end{align*}
In particular, when $|\calV| \to \infty$ is large, the second term vanishes.

\subsection{Putting it together}\label{subsec:puttingit}
As established previously:
$$
	\PP(v \text{ is not decidable at height h}) \le \PP( {\cal T}_h(v) \text{ not a tree} ) + \PP( \exists v \to v_1 \to ... \to v_h , e(v_i) = 0, i=1,...,h)
$$
The first term vanishes, and the second term is bounded as
\begin{align*}
	\PP( \exists v \to v_1 \to ... \to v_h & , \ell(v_i) = 0, i=1,...,h)  \le \sum_{v_1,...,v_h} \PP(v \to v_1 \to ... \to v_h , \ell(v_i) = 0, i=1,...,h) \\
	&\le n^h \PP( v \to v_1 \to ... \to v_h , \ell(v_i) = 0, i=1,...,h) \\
	&= n^h \PP(v \to v_1 \to ... \to v_h) \PP(\ell(v_i) = 0, i=1,...,h | v \to v_1 \to ... \to v_h) \\
	& = d^h \PP(\ell(v_i) = 0, i=1,...,h | v \to v_1 \to ... \to v_h) 
\end{align*}

For $i \in \{-1,+1\}$ define $\bar{\calG}_i$ the subgraph of nodes owned by player $i$ and whose priorities have the winning parity for player $i$. Define  $\bar{\calG} = \bar{\calG}_{-1} \cup \bar{\calG}_{+1}$ Any cycle in $\bar{\calG}$ is a self-winning cycle since all nodes have the winning parity for the corresponding player. Therefore, $\ell(v_i) = 0$ for $i=1,...,h$ implies that $v_i$ does not lead to a cycle in $\bar{\calG}$. In turn, this means that if we consider the exploration process of graph $\bar{\calG}$ starting at nodes $v,v_1,...,v_h$, then no cycle should appear.

It is noted that $\bar{\calG}$ is a graph where each node has a degree distribution Binomial$(d,{1 \over 4})$. Now consider the event $E = \{ v \to v_1 \to ... \to v_h \}$. Conditional to $E$, $\bar{\calG}$ is a graph where each node different from $v,v_1,...,v_{h-1}$ has a degree distribution Binomial$(d,{1 \over 4})$, and nodes $v,v_1,...,v_{h-1}$ have a degree distribution Binomial$(d-1,{1 \over 4})$. 

	Now consider the exploration process of graph $\bar{\calG}$ starting at nodes $v,v_1,...,v_h$. Since the number of successors of each node is lower bounded (in the strong stochastic sense) by Binomial $(d,{1 \over 4})$, we have that:
	$$
		\PP(e(v_i) = 0, i=1,...,h | E ) \le  \eta(d-1,{1 \over 4}(1-\varepsilon))^{h}  + \exp\left(- {d (\varepsilon |\calV|/d-1) \ln ((\varepsilon |\calV|/d-1)/(2 k)) \over 2 |\calV| \ln d}\right)
	$$
where we applied the result of the previous section. In turn 
\begin{align*}
	\lim\sup_{|\calV| \to \infty} \PP(v \text{ not decidable at height h}) &\le \lim\sup_{|\calV| \to \infty} \PP( {\cal T}_h \text{ not a tree} ) \\ &+  \lim\sup_{|\calV| \to \infty} \PP( \exists v \to v_1 \to ... \to v_h , e(v_i) = 0, i=1,...,h) \\
	&\le d^{h} \lim\sup_{|\calV| \to \infty} \PP(e(v_i) = 0, i=1,...,h | E ) \\
	&\le \Big[d \eta(d-1,{1 \over 4}(1-\varepsilon))\Big]^{h}.
\end{align*}
The above holds for all $\varepsilon > 0$ hence:
\begin{align*}
	\lim\sup_{|\calV| \to \infty} \PP(v \text{ not decidable at height h}) \le \Big[d \eta(d-1,{1 \over 4}) \Big]^{h}
\end{align*}	
Hence, with high probability, the {\tt SWCP} algorithm succeeds if 
$$
	d \eta\Big(d-1,{1 \over 4}\Big) < 1.
$$
This concludes the proof of theorem~\ref{th:main_result}.


\section{Numerical Experiments}

\begin{figure}[htbp!]
    \centering
    \includegraphics[scale = 0.75]{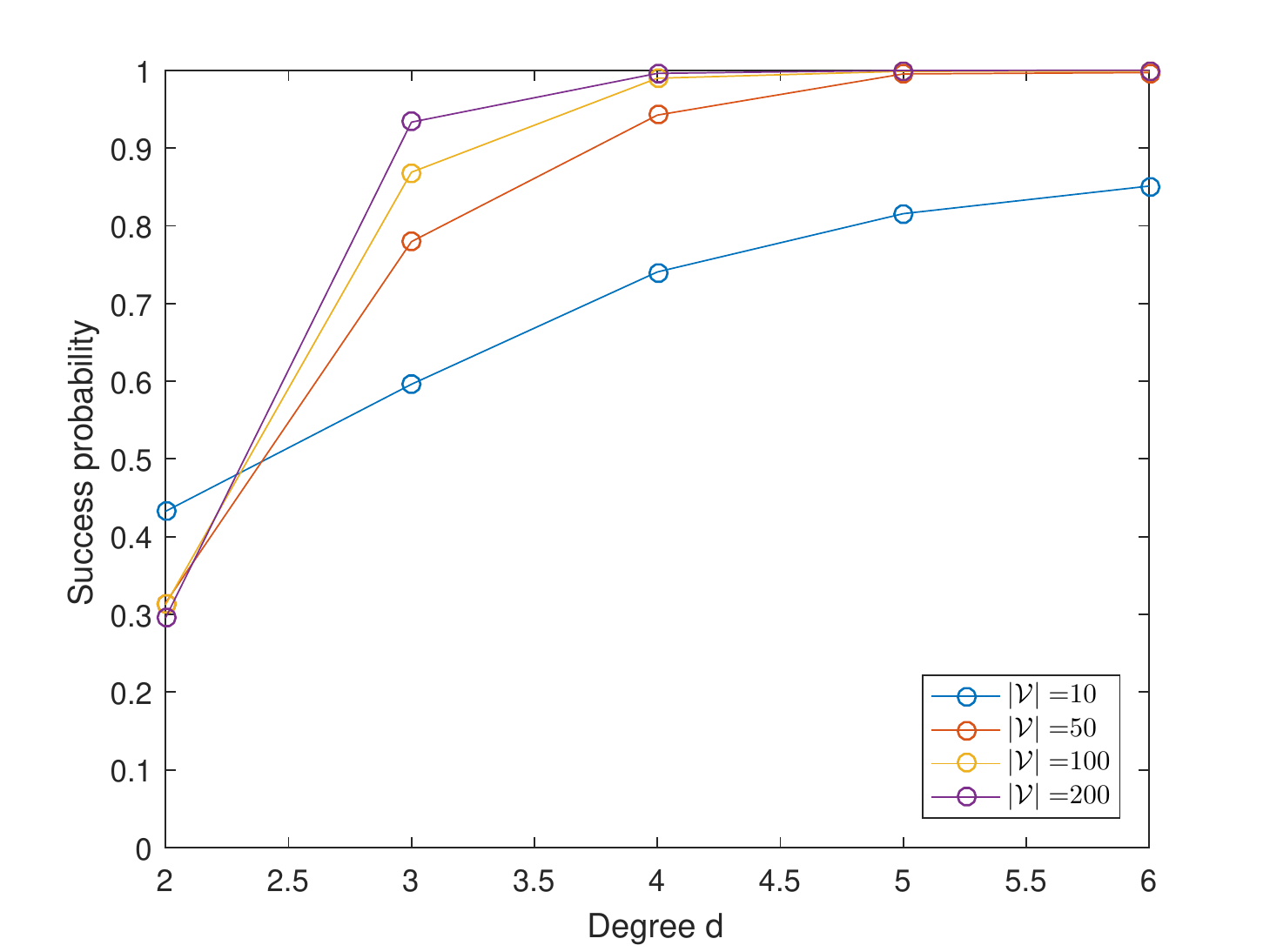}
    \caption{Probability of success vs $d$ and $|\calV|$}
    \label{fig:ProbabilityToSolve}
\end{figure}

\begin{figure}[htbp]
    \centering
    \includegraphics[scale = 0.75]{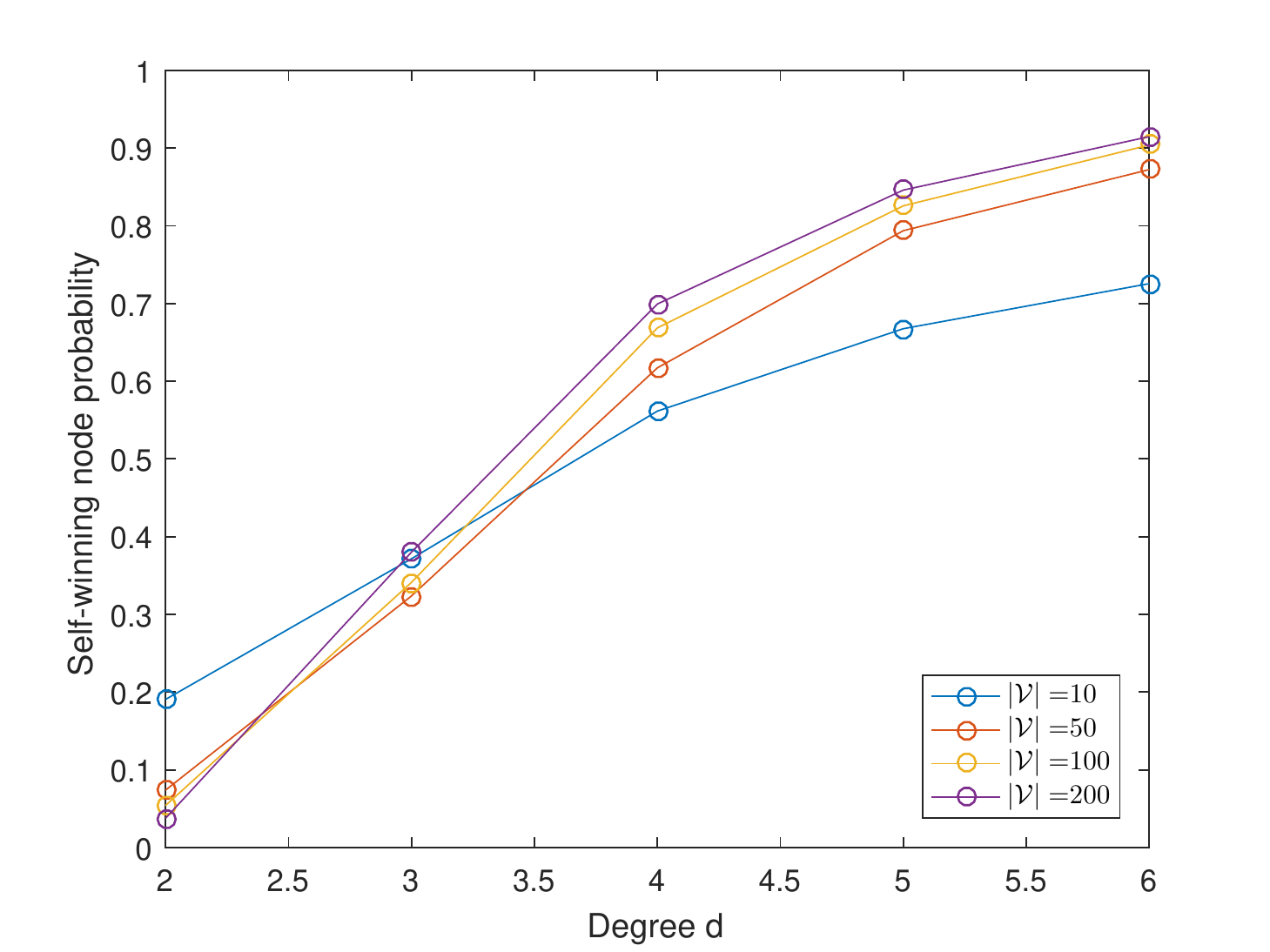}
    \caption{Proportion of self-winning nodes vs $d$ and $|\calV|$}
    \label{fig:ProbabilitySWC}
\end{figure}

\begin{figure}[htbp]
     \centering
     \includegraphics[scale = 0.75]{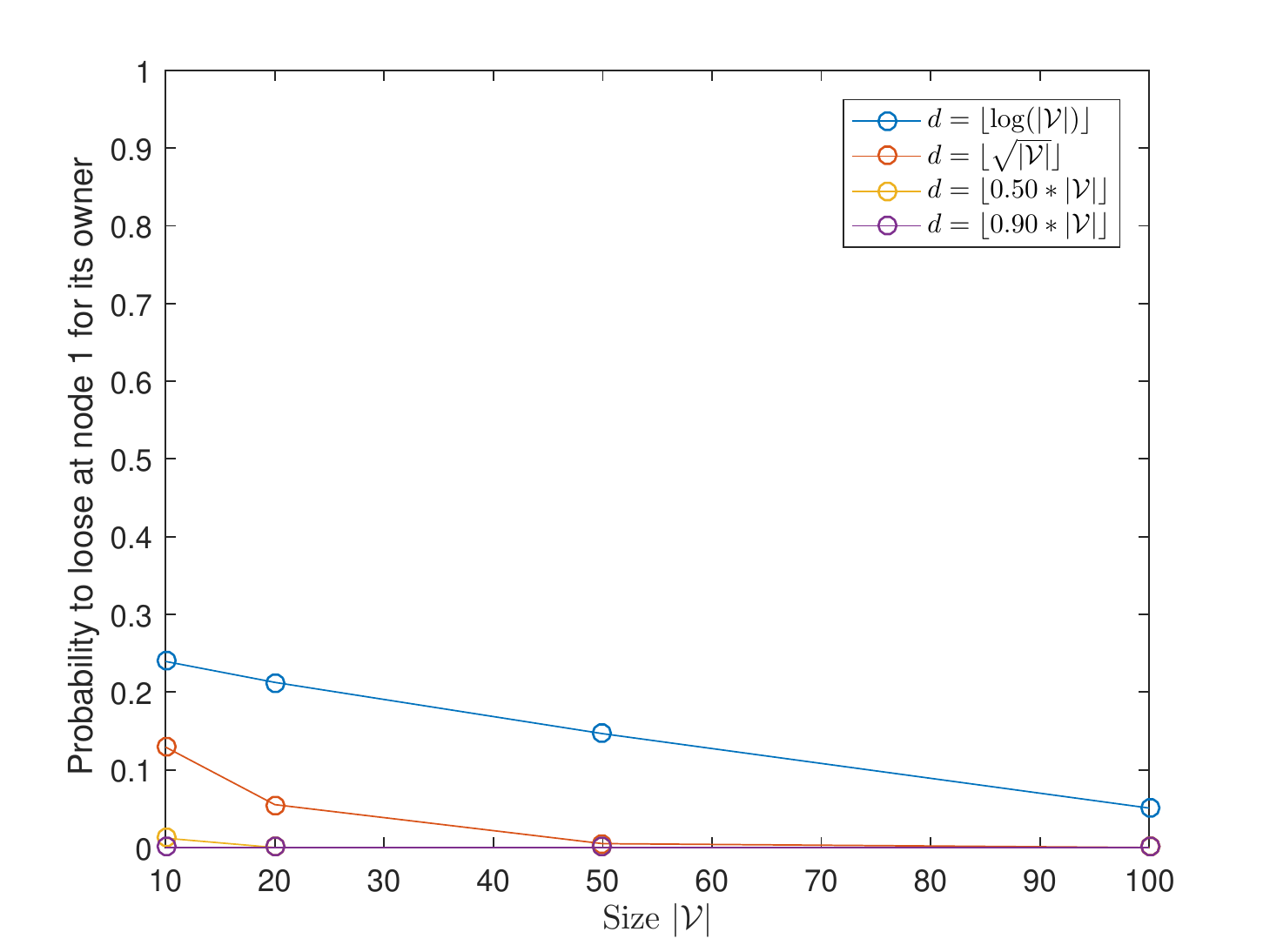}
     \caption{Probability that a node does not win for its owner in non sparse graphs.}
     \label{fig:probabilityNode1Looses}
\end{figure}

\begin{figure}[htbp]
     \centering
     \includegraphics[scale = 0.75]{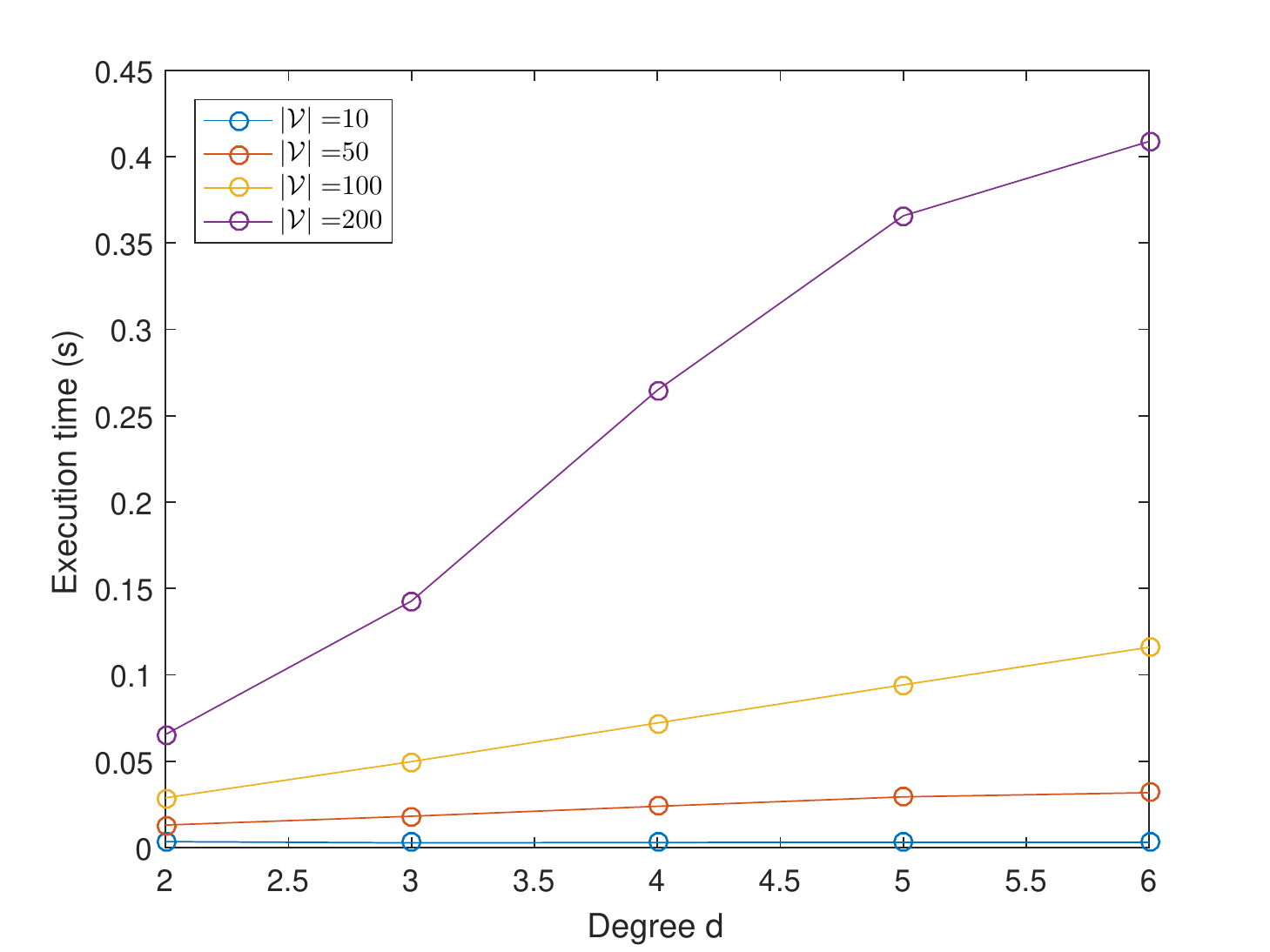}
     \caption{Average running time for {\tt SWCP}.}
     \label{fig:executionTime}
\end{figure}

We now illustrate our theoretical results using numerical experiments. Numerical experiments were computed with MatlabR2016a on a HP EliteBook 840G1 laptop with Inter Core i5-4300U CPU @1.90 GHz, 8Go RAM and Windows10.  For figures \ref{fig:ProbabilityToSolve}, \ref{fig:ProbabilitySWC},\ref{fig:probabilityNode1Looses} and \ref{fig:executionTime} we average the result over $200$ random instances of a parity game drawn as described in section~\ref{sec:main_result}. 

\begin{figure}[t!]
    \begin{subfigure}{0.50\textwidth}
        \includegraphics[width=0.9\linewidth]{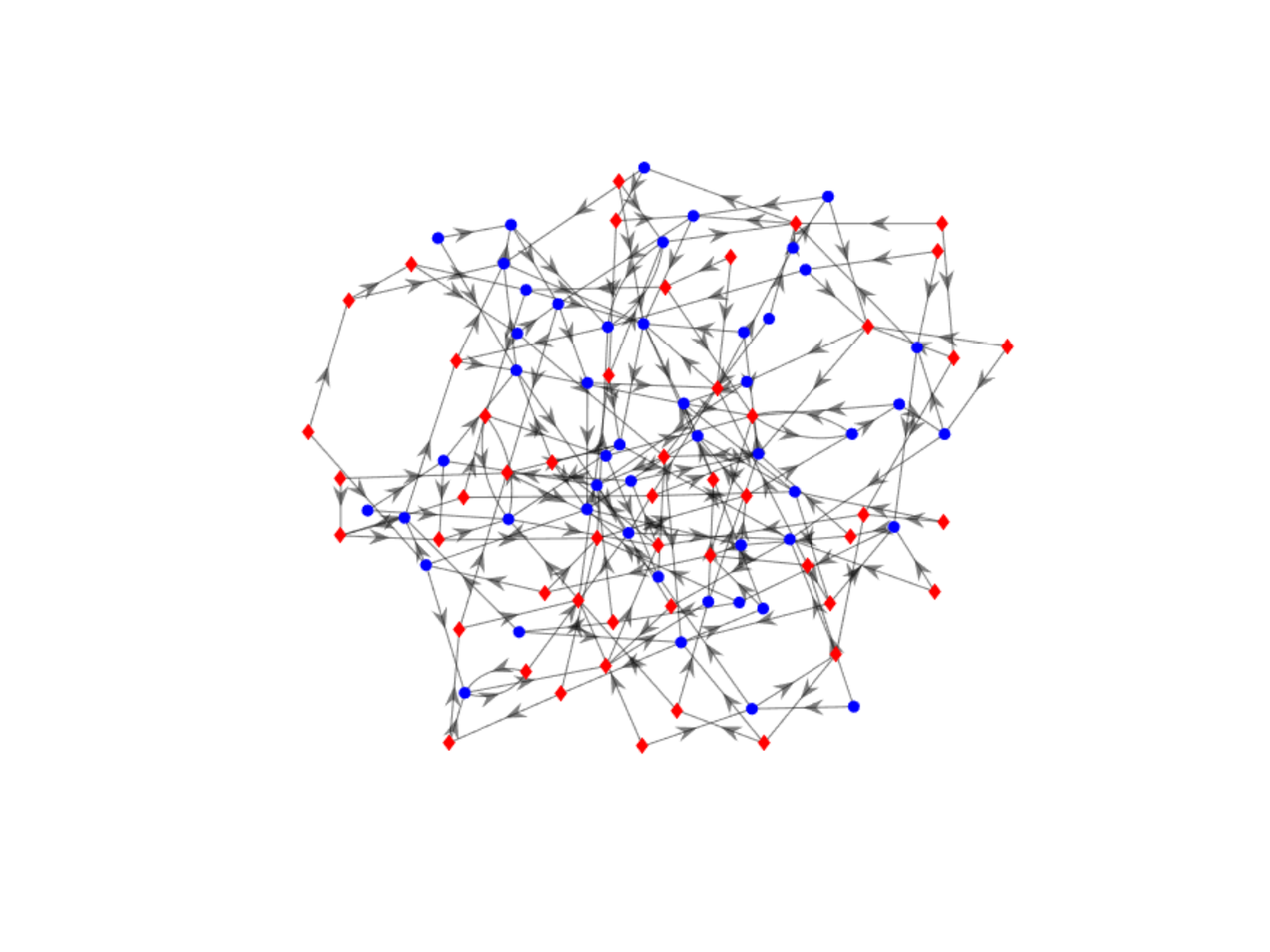} 
        \caption{$|\calV|=100$, $d=2$.}
        \label{fig:gameGraph_n100d2}
    \end{subfigure}
    ~
    \begin{subfigure}{0.50\textwidth}
        \includegraphics[width=0.9\linewidth]{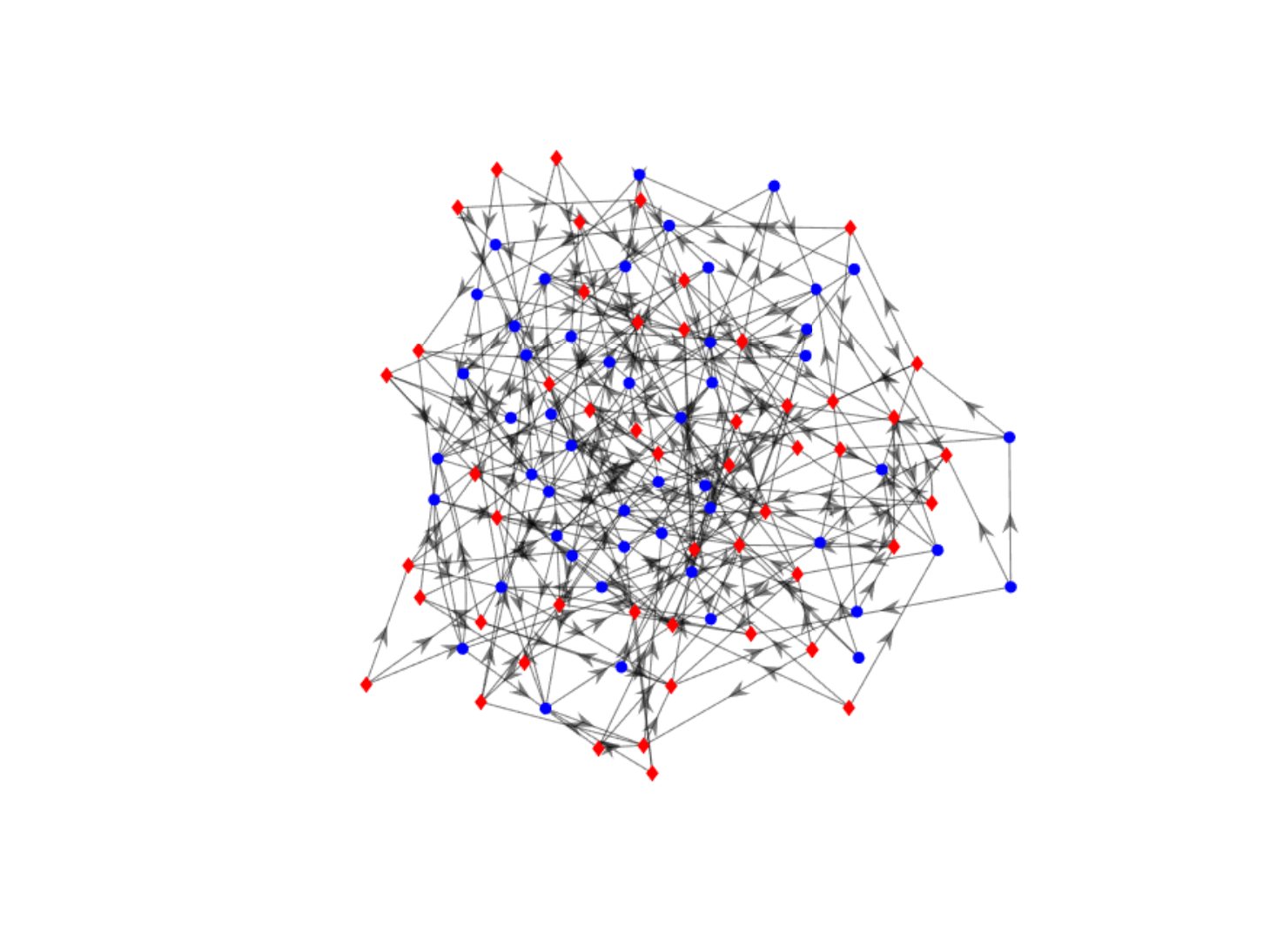}
        \caption{$|\calV|=100$, $d=3$.}
        \label{fig:gameGraph_n100d3}
    \end{subfigure}
    \begin{subfigure}{0.50\textwidth}
        \includegraphics[width=0.9\linewidth]{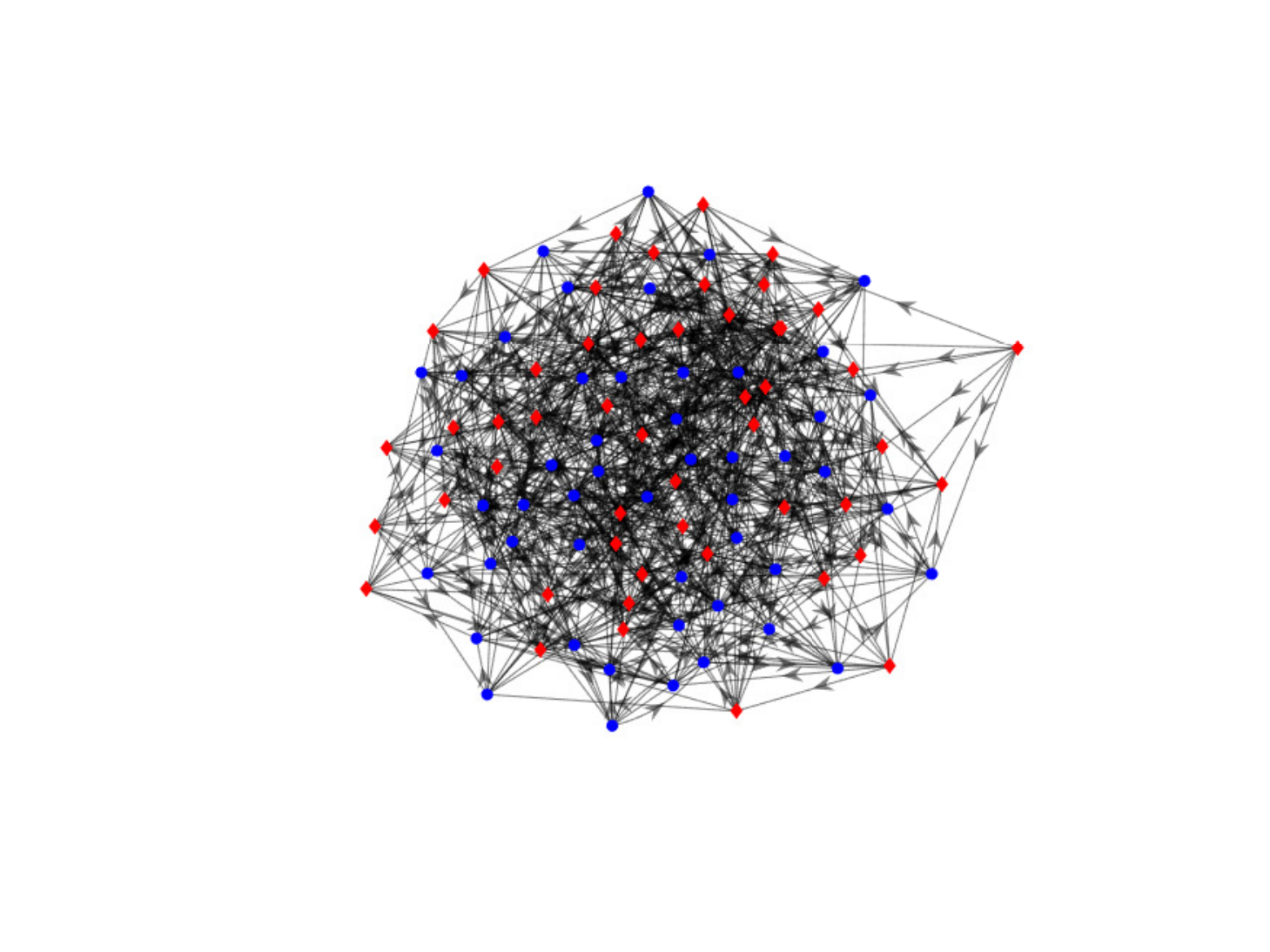} 
        \caption{$|\calV|=100$, $d=8$.}
        \label{fig:gameGraph_n100d8}
    \end{subfigure}
    ~
    \begin{subfigure}{0.50\textwidth}
        \includegraphics[width=0.9\linewidth]{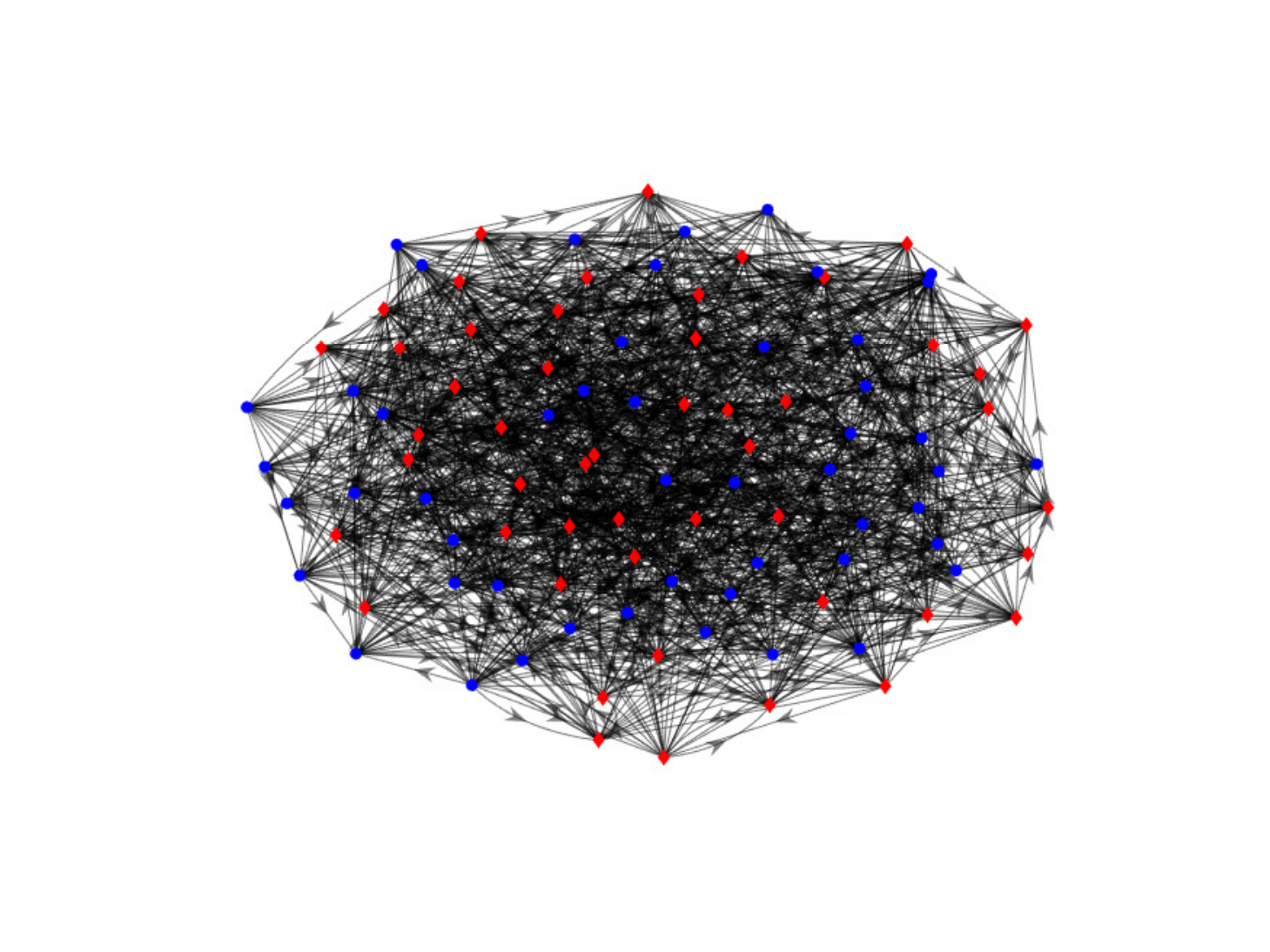}
        \caption{$|\calV|=100$, $d=15$.}
        \label{fig:gameGraph_n100d15}
    \end{subfigure}

    \caption{Game graph samples}
    \label{fig:gameGraphs2}
\end{figure}
Figure \ref{fig:gameGraphs2} displays samples of game graphs with $|\calV| = 100$ nodes and various degrees. For clarity we do not show the node priorities. Figure \ref{fig:ProbabilityToSolve} shows the probability that {\tt SWCP} outputs the value of a given node $b(v)$. As predicted by Theorem \ref{th:main_result} a phase transition occurs and above a threshold (which seems to be $d \ge 3$ on the figure), this probability goes to $1$ in the large graph regime. On the other hand, when $d=2$ the algorithm does not succeed with high probability.

Figure \ref{fig:ProbabilitySWC} shows the probability that a node is self-winning as a function of the degree $d$ and the number of nodes $|\calV|$, and this probability is, as expected, an increasing function of $d$.

Figure \ref{fig:probabilityNode1Looses} shows the probability that a given node is winning for the opponent in non-sparse graphs with degree equal to  $d=\floor{\ln |\calV| }, \floor{\sqrt{|\calV|}}, \floor{0.5 |\calV|},\floor{0.9 |\calV|}$. If the graph is sufficiently large and non-sparse, this probability tends to zero, so that then non sparse regime is easy to solve and the winner for each node is simply its owner as stated in Theorem~\ref{th:non_sparse}.

Figure \ref{fig:executionTime} shows the average execution time. 
For small graphs, the execution time is low and does not seem to depend on $d$. For larger graphs it does seem to grow linearly in the degree $d$, and quadratically in $|\calV|$, as predicted.


\section{Conclusion}

We have have studied random parity games and proposed  
the {\tt SWCP} algorithm which runs in time $O\Big(|\calV|^2 +  |\calV||\calE|\Big)$ and computes the value of the game w.h.p. for sparse games with large enough degree $d \ge d_P$. We have also shown that that non-sparse games and sparse games with $d=1$ can be solved w.h.p. in time $O(|\calV|)$ and $O(|\calV|^2)$ respectively. Finally we have formulated a challenging open problem, which is to determine whether or not one can solve sparse parity games in polynomial time in the $d=2$ case. 

\bibliographystyle{plain}
\bibliography{biblio}
	
\section{Appendix}

\subsection{Extended related work}\label{subsec:extended_related_work} 

\cite{Mcnaughton1993} studies Muller games and shows that the players have finite memory winning strategies, the existence of memoryless strategies for parity games and give an exponential recursive algorithm computing the winning positions and strategies. 

\cite{Zielonka1998} studies Muller games and parity games on finite and infinite graphs, showing a new inductive and constructive proof of the memoryless determinacy of parity games and an algorithm to compute the winning sets and winning strategies in the finite case. 

This algorithm is exponential \cite{Friedmann2011}, can be implemented in $O(|\calE|(|\calV|/c)^c)$ \cite{Jurdzinski2000}, performs very well in practice \cite{Friedmann2009}\cite{vanDijk2018} and keeps on inspiring research as \cite{Parys2019a}.

\cite{Jurdzinski2000} shows a progress measures lifting algorithm (see \cite{Jurdzinski2017} for an overview of progress measures for parity games) computing the winning sets and a winning strategy for player $Even$ with worst-case running in time $\calO\left(|\calE|\left(|\calV|/\floor{c/2}\right)^{\floor{c/2}}\right)$ and space $\calO(c |\calV|)$. 

\cite{Voge2000} shows a discrete strategy improvement algorithm computing winning strategies by using a local search technique and discrete vertex valuations to avoid past difficulties of other strategy improvement algorithms requiring high precision arithmetic (inefficient in practice).

\cite{Petersson2001} shows a randomized algorithm relying on reductions to mean and discounted payoff games and running in subexponential expected time $2^{O(c^{1/(1+2\epsilon)})}$ for a "large" number of colors $c = \Omega(|\calV|^{1/2+\epsilon})$, $0<\epsilon\leq 1/2$. 

\cite{Bjorklund2003} inspires from \cite{Ludwig1995} and proposes a randomized algorithm with complexity $$\min \left( O\left( |\calV|^3(\frac{|\calV|}{c}+1)^c \right), 2^{O(\sqrt{|\calV|\ln(|\calV|)})} \right)$$. If $c$ is small, the complexity is comparable to previous algorithms but the subexponential bound is an advantage if $c = \Omega (|\calV|^{1/2+\epsilon})$. Furthermore, subexponential algorithms were exponential for graphs with unbounded vertex out-degree, this drawback is solved. This complexity was also achieved by subexponential polynomial space deterministic algorithm \cite{Jurdzinski2008}\cite{Jurdzinski2006} running in $|\calV|^{O\sqrt{|\calV|/\ln |\calV|}}$ if the out-degree of all vertices is bounded, and $|\calV|^{\sqrt{|\calV|}}$ otherwise.

\cite{Schewe2007} shows a big-step algorithm combining the classical recursive techniques \cite{Mcnaughton1993}\cite{Zielonka1998} with progress measures lifting \cite{Jurdzinski2000} and dominion removal \cite{Jurdzinski2006}\cite{Jurdzinski2008}, improving the complexity to $O(|\calE||\calV|^{\frac{1}{3}c})$.  

\cite{Schewe2008} shows a new strategy improvement algorithm for parity and payoff games with optimal local strategy modifications. This algorithm improves previous strategy improvement algorithms by making optimal improvements, overcoming drawbacks due to the usual "update then evaluate" process.

The recent breakthrough came from \cite{Calude2017} proposing a quasi-polynomial algorithm deciding the winner in at most $O(|\calV|^{\ceil{\ln |\calV|}+6})$ and computing a memoryless winning strategy in $O(|\calV|^{7 + \ln c}\ln |\calV|)$. If $c<\ln |\calV|$, then the time complexity is $O(|\calV|^5)$, improved to $O(|\calE||\calV|^{2.55})$ in \cite{Gimbert2017}. This algorithm, relies on a polylogarithmic-space safety automaton (deciding the winner of a play) and its combination with the original parity game into an easily solvable safety game. 

\cite{Fearnley2019,Fearnley2017} shows a progress measure-based quasi-polynomial algorithm combining the succint coding from\cite{Calude2017} with progress measures and backward processing (allowing for some value iteration). For a fixed number of colours, the running time is quasi bi-linear $O(|\calE||\calV|(\ln|\calV|)^{c-1})$ , upper bounded by $O(|\calE|\ceil{1+c/\ln_2|\calV|}|\calV|^{\ln_2 e +\ln_2 \ceil{1+c/\ln_2(|\calV|)}})$ for a high number of priorities
and polynomial if $c=O(\ln |\calV_{even}|)$ where $\calV_{even}$ is the set of nodes with even priorities.

\cite{Lehtinen2018} shows a quasi-polynomial algorithm with complexity $|\calV|^{O(\ln |\calV|)}$ (see also \cite{Parys2019b} for an analysis of this work) using a nondeterministic safety automaton.

\cite{Jurdzinski2017} shows a quasi-polynomial time and near linear space algorithm.  It is shown that every progress measure on a finite game graph is equivalent to a succintly represented one, proving by the way new tree coding results and improving a bound on the number of lifts per vertex (see \cite{Jurdzinski2000}) by using the connection between a progress measure on a graph and vertices labelling by leaves of an ordered tree.

Denoting $\psi$ the smallest even number that is not smaller than the priority of any vertex, if $\psi \leq \ln |\calV|$, the polynomial upper bound is improved to $O(|\calE||\calV|^{2.38})$. 

For $\psi =\omega(\ln |\calV|)$, \cite{Jurdzinski2017} shows the $O(\psi |\calE| |\calV_{odd}|^{\ln_2(\psi/\ln \eta)+1.45})$ upper bound on the running time which is similar to the one obtained in \cite{Fearnley2019} when $\psi \geq \ln^2 |\calV_{odd}|$ and to be compared to $O(\psi |\calE| |\calV|^{\ln_2(\psi\ln_2 |\calV|) + 1.45})$ if $\psi = \Omega(\ln^2 |\calV_{odd}|)$ for Calude's algorithm as shown by \cite{Gimbert2017}.

More generally, \cite{Czerwinski2019} shows (see \cite{Parys2019a,Parys2019b,Jurdzinski2017} for overviews) a unified perspective on \cite{Calude2017,Jurdzinski2017,Lehtinen2018,Fearnley2017}, considering these works as instances of the separation approach, a technique relying on separating automata (nondeterministic in \cite{Lehtinen2018}, deterministic for others, these automata accepts plays consistent with winning memoryless strategies and rejects others with an exception for winning plays from non-memoryless strategies).

\cite{Parys2019a} shows a simple graph-dependant quasi-polynomial quadratic-space modification of Zielonka's algorithm.  
Despite of its simplicity, quasipolynomial complexity and expected good performances due to its foundation on the practically efficient Zielonka's algorithm, tests do not meet the expectations (particularly on random games and despite of good performances when the number of priorities is low). 

\cite{Parys2019b} studies Lehtinen's algorithm \cite{Lehtinen2018} and shows the conditions for nondeterministic separating automata to be used to solve parity games (defining the class of suitable-for-parity-games separators).
Lehtinen's algorithm complexity is improved from $|\calV|^{O( \ln \psi \,\ln |\calV|)}$ to $|\calV|^{O(\ln |\calV|)}$. 

From a stochastic games perspective, \cite{Chatterjee2004} considers a stochastic generalization of parity games with turn-based probabilistic transitions (some nodes belong to an additional player called "Random" selecting a successor uniformly at random.). They show a polynomial-time algorithm computing the winning sets in the single-player case (parity Markov decision process) and running in $O(c|\calE|^{3/2})$ ($c$ colours, $|\calE|$ edges), the existence of optimal memoryless strategies in the two-players case and that computing the values of the vertices is in NP $\cap$ co-NP. 

\cite{deAlfaro2004} studies infinite two-player games (simultaneous or turn-based, deterministic or probabilistic) with $\omega$-regular winning conditions.  It is shown that the maximal probability of winning is a fixed point of the quantitative game $\mu$-calculus. Winning strategies are characterized in terms of optimality and memory requirements. 

From a practical perspective, \cite{vanDijk2018} compares various algorithms (see \cite{Friedmann2009} for more tests), showing that Zielonka's algorithm \cite{Zielonka1998} and priority promotion \cite{Benerecetti2018} perform well, leaving room for new practically efficient quasi-polynomial algorithms. 

\subsection{Bipartite Parity Games}\label{subsec:bipartite_games}

By taking $\calE \subseteq \calV\times \calV$ we have not restricted to bipartite graphs ($\calE \subseteq \calV_E\times\calV_0 \times \calV \bigcup \calV_O\times\calV_E$), thus focusing the class of position-based parity games \cite{Calude2017}. 

Position-based (non-bipartite directed game graphs) and turn-based (bipartite directed game graphs) models can be translated into each others in polynomial time \cite{Calude2017}. 

The rationale for converting a position-based model to a turn-based one is the following:  for any edge $e=(v,v')$ of nodes belonging to the same player, delete $e$, introduce an intermediate node $w$ belonging to the opponent and reconnect with edges $(v,w)$ and $(w,v')$.
Node $w$ has in and out-degree one and is given a priority lower than $\max\{\Omega(v),\Omega(v')\}$ (so that the value of any self-winning cycle is not changed). 
In our case, the resulting game graph is not $d$-regular but it inherits the properties of existence of cycles of the original game graph (cycle length are doubled). The search for self-winning cycles and values propagation can either be done in the original subgraph or in the augmented one in the augmented subgraph.

\subsection{Beyond Parity Games}\label{subsec:beyond_parity_games}


Parity games are Muller games \cite{Mcnaughton1993}\cite{Grishpun2014} 
(used to solve to Church’s synthesis problem \cite{Buchi1969}\cite{Thomas2009}, the problem consists in finding a finite-state procedure turning an input sequence into an output sequence such that the pair satisfies a specification), 
the latter being regular games (including Muller, Rabin, Streett, Rabin chain, Parity and Büchi winning conditions) \cite{Gradel2002}\cite{Chatterjee2012}, 
included themselves in Borel games (Gale–Stewart games whose payoff sets are Borel sets) proven determined in the celebrated paper \cite{Martin1975}.
Parity games are remarkably fundamental since, for any regular game there exists a Muller game on the same game graph with the same winning region, and, for any Muller game there exists an equivalent (in deciding the winner)  parity game (see \cite{Gradel2002}).
The finite memory determinacy of Muller games was shown in \cite{Gurevitch1982} and the memoryless determinacy of parity games was shown in \cite{Mostowski1991} and \cite{Emerson1991}.
In addition to the latter connections, there are polynomial time reductions of parity games (see \cite{Petersson2001} for an overview) to other games as mean payoff games (the reduction is used to show that deciding the winner of a parity game is in UP $\cal$ Co-UP) \cite{Jurdzinski1998}, 
discounted games and simple stochastic games \cite{Zwick1996}. 

\subsection{Links with with $\mu$-calculus, mathematical logic and model checking} \label{subsec:mu_calculus}

See \cite{Gradel2002} for a survey and introduction monograph on automata, logics and infinite games.
 
$\mu$-calculus is a powerful logic for specifying and checking properties of transition systems.
It is one of the most important logics in model checking \cite{Baier2008}, subsuming other logics thanks to its expressiveness power, algorithmic properties and strong connections to games, particularly to parity games (see \cite{Bradfield2001}\cite{Bradfield2005}\cite{Bradfield2007} for introductions to $\mu$-calculus and links to games).
\cite{Emerson2001}\cite{Emerson1993} shows that the $\mu$-calculus model checking problem \cite{Emerson1986} (given a Kripke structure $M$, a state $s$ and a $\mu$-calculus formula $\phi$, $M,s\models \Phi$ ?) is equivalent to the non-emptiness problem of finite-state automata on infinite binary trees with a parity acceptance condition and polynomial time equivalent to deciding the winner of a parity game \cite{Emerson1993}\cite{Stirling1995}. 
Thus, determining the complexity of parity games would determine the complexity of $\mu$-calculus model checking, another major problem.

From an application perspective, model checking is particularly important in automated hardware and software verification and solving parity games is the central and most expensive step in many model checking, satisfiability checking and synthesis algorithms.

\subsection{Reminder about branching processes}\label{subsec:branching_reminder}

A branching process is a stochastic process described by the following evolution equations: $Z_0 = 1$ and
$$
	Z_{n+1} = \sum_{i=1}^{Z_n} X_{n,i}
$$
where $(X_{n,i})$ are i.i.d. drawn from some distribution called the offpsring distribution with mean $\mu$. When $\mu < 1$ the process is said to be subcritical and eventually gets extinct $\lim_{n \to \infty} \PP(Z_n = 0) = 1$. When $m > 1$ the process is said to be supercritical and the process either grows at an exponential rate or gets extinct with probability $\lim_{n \to \infty} \PP(Z_n = 0) = \eta < 1$. The extinction probability is the smallest $\eta$ verifying 
$$
 \EE(\eta^{X_{i}}) = \eta.
$$
In particular, for a branching process with offpsring distribution Binomial$(d,q)$ the extinction probability is the smallest solution to
$$
	(1-q + q \eta)^{d} = \eta 
$$
For more details on branching processes and random graphs the reader may consult~\cite{Hofstad2016}.

\subsection{Proof of Theorem~\ref{th:non_sparse}}

Recall that we write $\ell(v) = a(v)$ if there exists a path in subgraph $\calG_{a(v)}$ from $v$ to a self winning node and $\ell(v) = 0$ otherwise. Based on the same reasoning as the proof of Theorem~\ref{th:non_sparse}, we can prove that, for any $\varepsilon > 0$:
$$
		\PP(\ell(v)=0) \le  \eta(d-1,{1 \over 4}(1-\varepsilon))  + \exp\left(- {d (\varepsilon |\calV|/d-1) \ln ((\varepsilon |\calV|/d-1)/2) \over 2 |\calV| \ln d}\right)
	$$
Recall that $d = f(|\calV|) \to \infty$ when $|\calV| \to \infty$. Also we have $\eta(d-1,{1 \over 4}(1-\varepsilon)) \to 0$ when $d \to \infty$. Letting $|\calV|) \to \infty$ in the expression above yields the result:
$$
	\PP(\ell(v)=0 ) \to 0  \;,\; |\calV| \to \infty
$$
Since $\ell(v) \ne 0$ implies that $b(v) = a(v)$, this proves the result.

\subsection{Proof of proposition~\ref{prop:d2}}

Consider $d=2$. Consider node $v \in \calV$ and assume that $v$ is self-winning. This implies that $v$ in included in a cycle of subgraph $G_{a(v)}$ by definition. Consider $h \in \mathbb{N}$ and denote by $\calA_h$ and $\calB_h$ the events that $v$ is included in a cycle of $G_{a(v)}$ of length smaller or equal to $h$ (respectively) greater than $h$ so that
$$
 \PP(v \text{ is self-winning}) \le \PP(\calA_h) + \PP(\calB_h).
$$
The probability of $\calA_h$ is upper bounded by the expected number of cycles in $G_{a(v)}$ that contain $v$ using the classical argument:
$$
\PP(\calA_h) \le  \sum_{k=2}^h \Big({d \over 2 |\calV|}\Big)^k {(|\calV| - 1)! \over (|\calV| - k)!}
= \sum_{k=2}^h |\calV|^{-k} {(|\calV| - 1)! \over (|\calV| - k)!} \le {h \over |\calV|} \to 0 \;,\; |\calV| \to \infty
$$
The probability of $\calB_h$ is upper bounded by the probability that there exists at least $h$ nodes that can be reached from $v$, in turns this implies that $S_h > 0$ where $S$ is the exploration process (see subsection \ref{subsec:preliminary}) of subgraph $\calG_{a(v)}$ starting at node $v$. In subgraph $\calG_{a(v)}$, each node has a number of nodes distributed as Binomial($2$,${1 \over 2}$).

By a similar argument as our previous analysis, $\PP(S_h > 0)$ is upper bounded by the probability that a branching process with offspring distribution Binomial($2$,${1 \over 2}$) has a total progeny larger than $h$. Since this branching process is critical, and the progeny distribution is not constant, it undergoes extinction almost surely, which proves that 
$\PP(S_h > 0) \to 0$ when $h \to \infty$. 

This proves that for all $h$: 
$$
\PP(v \text{ is self-winning}) \le {h \over |\calV|} + \PP(S_h > 0)  
$$
and by setting $h = \sqrt{|\calV|}$ and letting $|\calV| \to \infty$ we get that
$$
\PP(v \text{ is self-winning}) \to 0 \;,\; |\calV| \to \infty.
$$
	
\end{document}